\documentclass[11pt]{article}
\usepackage{amsmath,amssymb}
\usepackage{color}

\makeatletter
\@addtoreset{equation}{section}
\makeatother

\def\ben{\begin{equation}}
\def\een{\end{equation}}

\let\a=\alpha \let\b=\beta  \let\d=\delta 
    
 \let\m=\mu \let\n=\nu  \let\p=\pi

\let\C=\Chi

\def\nn{\nonumber} \def\bd{\begin{document}} \def\ed{\end{document}}
\def\ds{\documentstyle} \let\fr=\frac \let\bl=\bigl \let\br=\bigr
\let\Br=\Bigr \let\Bl=\Bigl
\let\bm=\bibitem
\let\na=\nabla
\let\pa=\partial \let\ov=\overline
\newcommand{\be}{\begin{equation}}
\newcommand{\ee}{\end{equation}}
\def\ba{\begin{array}}
\def\ea{\end{array}}
\def\ft#1#2{{\textstyle{\frac{\scriptstyle #1}{\scriptstyle #2} } }}
\def\fft#1#2{{\frac{#1}{#2}}}
\def\del{\partial}
\def\vp{\varphi}
\def\sst#1{{\scriptscriptstyle #1}}
\def\oneone{\rlap 1\mkern4mu{\rm l}}
\def\td{\tilde}
\def\wtd{\widetilde}
\def\ie{{\it i.e.\ }}
\def\dalemb#1#2{{\vbox{\hrule height .#2pt
        \hbox{\vrule width.#2pt height#1pt \kern#1pt
                \vrule width.#2pt}
        \hrule height.#2pt}}}
\def\square{\mathord{\dalemb{6.8}{7}\hbox{\hskip1pt}}}
\newcommand{\ho}[1]{$\, ^{#1}$}
\newcommand{\hoch}[1]{$\, ^{#1}$}
\newcommand{\bea}{\setlength\arraycolsep{2pt} \begin{eqnarray}}
\newcommand{\eea}{\end{eqnarray}}
\newcommand{\ra}{\rightarrow}
\newcommand{\lra}{\longrightarrow}
\newcommand{\Lra}{\Leftrightarrow}
\newcommand{\bp}{\tilde \beta^\prime}
\newcommand{\tr}{{\rm tr} }
\newcommand{\Tr}{{\rm Tr} }
\def\0{{\sst{(0)}}}
\def\1{{\sst{(1)}}}
\def\2{{\sst{(2)}}}
\def\3{{\sst{(3)}}}
\def\4{{\sst{(4)}}}
\def\5{{\sst{(5)}}}
\def\6{{\sst{(6)}}}
\def\7{{\sst{(7)}}}
\def\8{{\sst{(8)}}}
\def\m{{\sst{(m)}}}
\def\n{{\sst{(n)}}}
\def\cA{{{\cal A}}}
\def\cB{{{\cal B}}}
\def\cF{{{\cal F}}}
\def\cG{{{\cal G}}}
\def\cH{{{\cal H}}}
\def\tV{\widetilde V}
\def\tW{\widetilde W}
\def\tH{\widetilde H}
\def\tE{\widetilde E}
\def\tF{\widetilde F}
\def\tA{\widetilde A}
\def\im{{{\rm i}}}
\def\tY{{{\wtd Y}}}
\def\ep{{\epsilon}}
\def\vep{{\varepsilon}}
\def\bD{{{\bar D}}}
\def\R{{{\mathbb R}}}
\def\C{{{\mathbb C}}}
\def\H{{{\mathbb H}}}
\def\CP{{{\mathbb C}{\mathbb P}}}
\def\RP{{{\mathbb R}{\mathbb P}}}
\def\Z{{{\mathbb Z}}}
\def\bA{{{\mathbb A}}}
\def\bB{{{\mathbb B}}}
\def\bC{{{\mathbb C}}}
\def\bD{{{\mathbb D}}}
\def\bE{{{\mathbb E}}}
\def\bZ{{{\mathbb Z}}}
\def\Re{{{\frak{Re}}}}
\def\Im{{{\frak{Im}}}}
\def\cosec{{\,\hbox{cosec}\,}}
\def\Gm{{\Gamma_{\!\! -}}}
\def\Gp{{\Gamma_{\!\! +}}}
\def\stan{{standard }}
\def\nonstan{{supernumerary }}
\def\p{{\partial}}
\def\kdel#1{{\fft{\del}{\del#1}}}

\def\bog{{Bogomolny }}
\def\om{{\omega}}

\newcommand{\nnr}{\nonumber \\}
\newcommand{\pd}{\partial}
\newcommand{\ud}{\textrm{d}}
\newcommand{\dTH}{T^{\prime \, 0}_\textrm{H}}
\newcommand{\dOi}{\Omega^{\prime \, 0}_i}

\newcommand{\tamphys}{\it George and Cynthia Woods Mitchell  Institute
for Fundamental Physics and Astronomy,\\
Texas A\&M University, College Station, TX 77843-4242, USA}

\newcommand{\auth}{
Zhao-Long Wang\hoch{\dagger} and H. L\"u\hoch{\star}
}

\begin{document}

\begin{flushright}
\hfill{
MIFP-09-26\ \ \ \ USTC-ICTS-09-09\ \ \ \  }\\
\end{flushright}

\begin{center}
{\large {\bf Most General Spherically Symmetric M2-branes
and Type IIB Strings}}

\vspace{15pt}
\auth

\vspace{10pt}
\hoch{\dagger}{\it Interdisciplinary Center for Theoretical Study,\\
University of Science and Technology of China, Hefei, Anhui
230026, China}

\vspace{10pt}

\hoch{\star}{\it 
China Institute for Advanced Study,\\
China Economics and Management Academy,\\
Central University of Finance and Economics, Beijing, 100081, China
}

\vspace{30pt}

\underline{ABSTRACT}
\end{center}

We obtain the most general spherically symmetric M2-branes and type
IIB strings, with $\R^{1,2}\times SO(8)$ and $\R^{1,1}\times SO(8)$
isometries, respectively.  We find that there are 12 different
classes of M2-branes, and we study their curvature properties.  In
particular, we obtain new smooth M2-brane wormholes that connect two
asymptotic regions: one is flat and the other can be either flat or
AdS$_4\times S^7$.  We find that these wormholes are traversable with
certain timelike trajectories.  We also obtain the most general
Ricci-flat solutions in five dimensions with $\R^{1,1}\times SO(3)$
isometries.

\vspace{15pt}

\thispagestyle{empty}

\pagebreak
\setcounter{page}{1}

\tableofcontents

\addtocontents{toc}{\protect\setcounter{tocdepth}{2}}


\vspace{30pt}

\section{Introduction}

    The higher dimensionality and the inclusion of higher form fields
in string and M theories allow a variety of $p$-brane solitons in the
spectrum \cite{dkl}.  These objects are generalizations of charged
black holes and have been playing an important role in understanding
the nonperturbative aspects of the string theories.  A $p$-brane
soliton is defined to be a spacetime configuration that splits the
$D$-dimensional spacetime into the world volume of $(p+1)$ dimensions
and the transverse space of $(D-p-1)$ dimensions.  The
electrically charged $p$-brane is supported by a $(p+1)$-form
potential, while the magnetically charged one is by a $(D-p-1)$-form.
The simplest type of $p$-branes has the Minkowski world volume and
spherically symmetric transverse space.  A classification of such
$p$-branes in maximal supergravities was given in \cite{lpsoliton}.
Lower-dimensional $p$-branes can be lifted to higher dimensions and
become intersecting M2-branes or type II strings \cite{aat,kt}.
Classifications of brane intersections in maximal supergravities were
given in \cite{lptx,bbj}.

      In this paper, we consider a more general class of $p$-brane
solutions, focusing on M2-branes and type IIB strings.  While we
still retain the spherical symmetry in the transverse space for
simplicity, we shall relax the condition for the world volume from
being Minkowskian to having instead the $\R^{1,p}$ isometry.  A simple
example is a $p$-brane with gravitational a pp-wave propagating in the
world volume.  Such a $p$-brane has the isometry of $\R^{1,p}\times
SO(D-p-1)$.  The metric takes the form
\be
ds^2_D = r^2 d\Omega_{D-p-2}^2 + \fft{dr^2}{f} + \sum_{\mu,\nu=0}^p
g_{\mu\nu} dz^\mu dz^\nu\,,
\ee
where $f$ and $g_{\mu\nu}$ depend only on the radial variable $r$ of
the transverse space.  The world volume coordinates are $z^\mu$, with
$z^0=t$ denoting the time direction.  The way to obtain such a
solution is to observe that one can perform Kaluza-Klein reduction on
the coordinates $z^\mu$, on which no field depends.  The resulting
solution in $(D-p-1)$ dimensions is an instanton supported by a
certain scalar coset $G/H$.  In the case of pure gravity, the scalar
coset is $\fft{GL(p+1, \R)}{SO(1,p)}$, for which the most general
solution was obtained in \cite{wjl}.  (See also \cite{bcptv,crtv}.)

    In section 2, we consider the reduction of M2-branes on the
world volume.  The Kaluza-Klein reduction of eleven-dimensional
supergravity on $\R^{1,2}$ gives rise to a scalar coset of
$\fft{SL(3,\R)}{SO(1,2)}\times\fft{SL(2,\R)}{SO(1,1)}$
\cite{euclid,hj}.  This enables us to make use of the results in
\cite{wjl} and obtain the most general M2-brane solutions.  We find
that there are twelve classes of solutions, including the previously
known extremal \cite{dust} or nonextremal M2-branes \cite{dlp,ct}.
We also find new smooth M2-brane wormhole solutions.  We study the
curvature properties of these solutions.  In particular, in section 3,
we focus on the wormholes and demonstrate that they are not
traversable geodesically but traversable with certain timelike
trajectories, as in the case of the previously known Ricci-flat
wormholes \cite{cd,aac,lmflat,wjl}.

    The $\fft{SL(3,\R)}{SO(1,2)}\times\fft{SL(2,\R)}{SO(1,1)}$ scalar
coset can also be obtained from the reduction of type IIB supergravity
on $\R^{1,1}$.  This enables us to lift the instanton solutions back
to $D=10$ and obtain the most general type IIB string solutions.  We
do this in section 4. It is worth pointing out that, in the case of M
theory on $\R^{1,2}$, the $SL(3,\R)$ global symmetry is part of the
general coordinate transformation, and hence we can use it to mod out
the equivalent solutions related by coordinate transformation.  In the
case of type IIB supergravity on $\R^{1,1}$, it is the $SL(2,\R)$
factor that is part of the general coordinate transformation.  Thus
the same instanton solution in $D=8$ can lead to very different
lifting to M2-branes or type IIB strings.

   In section 5, we further consider five-dimensional pure gravity on
$\R^{1,1}$, which becomes the scalar coset of $\fft{SL(3,\R)
}{SO(1,2)}$.  This enables us to obtain the most general Ricci-flat
solutions in five dimensions with $\R^{1,1}\times SO(3)$ isometry.

     We conclude the paper in section 6.

\section{General Spherically symmetric M2-branes}

     The world volume of M2-branes has three dimensions, including one
time and two spatial directions.  Performing Kaluza-Klein reduction on
the world volume gives rise to instanton solutions in $D=8$ Euclidean
maximal supergravity.  In this section, we review the scalar sector in
$D=8$ and obtain the most general spherically symmetric instanton
solutions supported by the scalars.  Lifting the solutions back to
$D=11$ gives rise to the most general spherically symmetric M2-branes,
with $\R^{1,2}\times SO(8)$ isometries.  We then study the curvature
properties of these solutions.

\subsection{$D=8$ Euclidean maximal supergravity: the scalar sector}
\label{reductionansatz}

     Eight-dimensional Euclidean maximal supergravity obtained from
$\R^{1,2}$ reduction of eleven-dimensional supergravity was discussed
in \cite{euclid,hj}. Here we shall focus on the scalar sector.  The
bosonic sector of eleven-dimensional supergravity consists of the
metric and a 3-form tensor field $A_\3$.  The action is given by
\cite{cjs11}
\be
S_{11} = \int d^{11} x \sqrt{-G} \left(R - \fft1{48} F_\4^2\right) -
\fft16 \int A_\3\wedge F_\4\wedge F_\4\,,
\ee
where $F_\4 = dA_\3$.  There are a total of seven scalars in $D=8$,
six of which come from the metric, whose reduction ansatz is given by
\bea
 ds^2_{11}&=&e^{-\fft13\Phi}ds_8^2 +  e^{\fft23\Phi}dz^{\rm T}
{\cal M}\,dz\cr
&=&e^{-\fft13\Phi}ds_8^2+ e^{\fft23\Phi}\left(
 e^{\Phi_1+\fft{1}{\sqrt3}\Phi_2}(dz_1+\chi_{12}dz_2+\chi_{10}dt)^2
\right.\cr
&&\left. ~~~~~~~~~~~~~~~~~~~~~~~~~
 +e^{-\Phi_1+\fft{1}{\sqrt3}\Phi_2}(dz_2+\chi_{20}dt)^2
 -e^{-\fft{2}{\sqrt3}\Phi_2}dt^2\right)\,,
 \eea
where $z=(z_1,z_2, t)$ and the matrix $\cal M$ in the metric involves
five scalars, parameterizing the $\fft{SL(3,\R)}{SO(1,2)}$ scalar
coset, namely,
{\fontsize{10 pt}{\baselineskip}\selectfont
\bea
 \cal M=\begin{pmatrix} \label{M}
e^{\Phi_1+\fft{\Phi_2}{\sqrt3}}\chi_{10}^2
-e^{-\fft{2\Phi_2}{\sqrt3}}+e^{-\Phi_1+\fft{\Phi_2}{\sqrt3}}
\chi_{20}^2\,\,\,
 &e^{\Phi_1+\fft{\Phi_2}{\sqrt3}}\chi_{10} &\,\,\,
 e^{\Phi_1+\fft{\Phi_2}{\sqrt3}}\chi_{12}\chi_{10}+
e^{-\Phi_1+\fft{\Phi_2}{\sqrt3}}\chi_{20}\\
 e^{\Phi_1+\fft{1}{\sqrt3}\Phi_2}\chi_{10}
& e^{\Phi_1+\fft{1}{\sqrt3}\Phi_2}
& e^{\Phi_1+\fft{1}{\sqrt3}\Phi_2}\chi_{12} \\
 e^{\Phi_1+\fft{1}{\sqrt3}\Phi_2}\chi_{12}\chi_{10}+
e^{-\Phi_1+\fft{1}{\sqrt3}\Phi_2}\chi_{20} & e^{\Phi_1+
\fft{1}{\sqrt3}\Phi_2}\chi_{12} & e^{-\Phi_1+\fft{1}{\sqrt3}\Phi_2}+
e^{\Phi_1+\fft{1}{\sqrt3}\Phi_2}\chi_{12}^2
 \end{pmatrix}.\nn\\
\eea}
The parametrization of such a scalar coset was discussed in detail
in \cite{euclid,wjl}.  The reduction of the 3-form tensor field
$A_\3$ on $\R^{1,2}$ gives rise to one axionic scalar $\chi$, given
by
\be
A_\3 = \chi\, dt\wedge dz_1\wedge dz_2\,.
\ee
This axion and the breathing mode $\Phi$ in the metric form
an $\fft{SL(2,\R)}{SO(1,1)}$ scalar coset with
 \bea \tilde {\cal M}=\begin{pmatrix}
 -e^{\Phi}+e^{-\Phi}\chi^2 &\quad e^{-\Phi}\chi\\
 e^{-\Phi}\chi &\quad  e^{-\Phi}
 \end{pmatrix}.
 \eea
The Lagrangian of the scalar sector, coupled to gravity, of
Euclidean maximal supergravity in $D=8$ is then given by
\bea
 {\cal L}_8&=&\sqrt{g} \Big(R + \ft14 {\rm tr} (\partial_\mu {\cal
 M}^{-1}\partial^\mu{\cal M}) + \ft14 {\rm tr} (\partial_\mu
{\tilde{\cal M}}^{-1}\partial^\mu{\tilde{\cal M}})\Big)\cr
 &=&\sqrt{g} \Big(R-\ft12(\partial\Phi)^2
 +\ft12e^{-2\Phi}(\partial\chi)^2-\ft12(\partial\Phi_1)^2
-\ft12(\partial\Phi_2)^2\label{d8lag1}\\
&&~~~~-\ft12e^{2\Phi_1}(\partial\chi_{12})^2+
 \ft12e^{\Phi_1+\sqrt3\Phi_2}(\partial\chi_{10}-
\chi_{20}\partial\chi_{12})^2+\ft12e^{-\Phi_1+\sqrt3\Phi_2}
(\partial\chi_{20})^2\Big)\,.\nn
 \eea

\subsection{Spherically symmetric solution}

We now consider spherically symmetric instanton solutions to the
Lagrangian (\ref{d8lag1}) in eight dimensions.  The metric ansatz is
given by
\be
ds_8^2 = \fft1{f(r)} dr^2 + r^2 d\Omega_7^2\,,\label{ds8}
\ee
with all of the scalars depending only on the radial coordinate $r$.  The
equations of motion of such an ansatz for general scalar cosets were
discussed in \cite{wjl}.  The Einstein equations in the directions of
the foliating sphere $S^7$ imply that
\be
12 (1- f) - r f' =0\,,
\ee
where a prime denotes a derivative with respect to $r$.  Thus we
have
\be
f=1 - \fft{a}{r^{12}}\,.\label{genfsol}
\ee
The Einstein equation associated with $R_{rr}$ implies that
\be
-\fft{7f'}{2r\,f} + \ft14 \tr({\cal M}^{-1'} {\cal M}') +
\ft14 \tr({\cal \tilde M}^{-1'} {\cal \tilde M}')=0\,.
\label{slnrrr}
\ee
The scalar equations of motion are given by
\be ({\cal M}^{-1} \dot {\cal M})\,\dot{} =0~,\qquad
{(\tilde{\cal M}^{-1} \dot
 {\tilde{\cal M}})\,\dot{}} =0  \label{slnrscalareq}
 \ee
where a dot denotes a derivative with respect to $\rho$, defined by
\be
d\rho=\fft{dr}{r^{7}\sqrt{f}}\,. \ee
Thus, we have
\be
\rho=-\fft{1}{6\sqrt a}\arcsin\left(\frac{\sqrt a}{r^6}\right)\,.
\ee
The second-order differential equations (\ref{slnrscalareq}) for the
scalars can be easily integrated to give rise to a set of first-order
equations, given by
 \be
\label{scalareom} {\cal M}^{-1} \dot {\cal M} = {\cal C}\,,
\qquad\tilde{\cal M}^{-1} {\dot{\tilde {\cal M}}} = \tilde{\cal C}\,,
\ee
where ${\cal C}$ and $\tilde{\cal C}$ are Lie-algebra valued constant
matrices. Both ${\cal C}$ and $\tilde{\cal C}$ are traceless since
${\det}{\cal M}={\det}{\tilde {\cal M}}=-1$.  Substituting this and
(\ref{genfsol}) into (\ref{slnrrr}), we have the following constraint
on these constant matrices, namely
\be
\label{Hamiltonian} {\cal I}\equiv -\ft12 {\rm tr} ({\cal
 C}^2)-\ft12 {\rm tr} ({\tilde {\cal C}}^2) = 2(D-1)(D-2) a=84a\,.
 \ee
Thus we see that the solutions are completely determined by the
constant matrices ${\cal C}$ and $\tilde {\cal C}$.  Such matrices for
the $SL(n,\R)/SO(1,n-1)$ scalar cosets were classified in
\cite{wjl,bcptv} that would yield different classes of
solutions. If we turn off the scalars in $\tilde {\cal M}$, the
solutions are then Ricci-flat,  and fully classified in
\cite{wjl}, including new smooth wormholes and new tachyon waves.
If instead we turn off the scalars in ${\cal M}$,  we may obtain
supersymmetric \cite{dust} and non-supersymmetric M2-branes as in
\cite{lpxtoda,lu&roy}.

  We now consider the classifications of ${\cal C}$ and $\tilde {\cal
C}$ and hence ${\cal M}$ and $\tilde {\cal M}$ in detail.  We begin
with ${\cal C}$ and ${\cal M}$, which were discussed in \cite{wjl}.
Here we shall just present the results.

{\bf Class I:} The constant matrix ${\cal C}$ has a pair of complex
eigenvalues. It is isomorphic to
\be
{\cal C} = \begin{pmatrix} \alpha & -\beta & 0\cr \beta & \alpha
 & 0\cr 0 & 0 & -2\alpha \end{pmatrix}\,.
 \ee
It follows that
\be
{\cal M} = \begin{pmatrix} -e^{\alpha\rho}\cos{\beta\rho} &
 e^{\alpha\rho}\sin{\beta\rho} & 0\cr e^{\alpha\rho}\sin{\beta\rho}&
 e^{\alpha\rho}\cos{\beta\rho} & 0\cr 0 & 0 & e^{-2\alpha\rho}
 \end{pmatrix}\,.
 \ee

{\bf Class II:} The matrix ${\cal C}$ has three real eigenvalues, with
one timelike and two spacelike eigenvectors. It is isomorphic to
 \be {\cal C} = \begin{pmatrix} \alpha_1 & 0 & 0\cr 0 & \alpha_2 &
 0\cr 0 & 0 & -(\alpha_1+\alpha_2)
 \end{pmatrix}\,.
 \ee
Thus we have
 \be {\cal M} = \begin{pmatrix} -e^{\alpha_1\rho} & 0 & 0\cr 0&
 e^{\alpha_2\rho} & 0\cr 0 & 0 & e^{-(\alpha_1+\alpha_2)\rho}
 \end{pmatrix}\,.
 \ee

    The matrix ${\cal C}$ can also have three real eigenvalues, but
with degenerate eigenspace. There are two inequivalent cases, leading
to the follow two classes.

{\bf Class III:} ${\cal C}$ is rank 2 and there is a twofold eigenvalue
with a 1-dimension eigenspace. Such a ${\cal C}$ is isomorphic to
 \be {\cal C} = \begin{pmatrix} \tilde Q+\a & \tilde Q & 0\cr
-\tilde Q & -\tilde Q+\a & 0 \cr 0
 & 0 &-2\a
 \end{pmatrix}\,.\label{c3case2}
 \ee
Note that we can fix $\tilde Q=\pm1$ by an $O(1,1)$ boost. The corresponding
matrix ${\cal M}$ is given by
 \be {\cal M} = \begin{pmatrix} -e^{\a\rho}(1+\tilde Q\rho) &
 -e^{\a\rho}\tilde Q\rho & 0\cr -e^{\a\rho}\tilde Q\rho& e^{\a\rho}
(1-\tilde Q\rho) & 0\cr 0 & 0 & e^{-2\a\rho}\end{pmatrix}\,.
 \ee

{\bf Class IV:} ${\cal C}$ is rank 2 and all of its eigenvalues are
zero. Such a ${\cal C}$ is isomorphic to
 \be {\cal C} = \begin{pmatrix}
 \cos\beta & \cos\beta & -\ft12 \sin\beta \\
 -\cos\beta & -\cos\beta &   \ft12\sin\beta\\
 \ft12\sin\beta & \ft12\sin\beta & 0\\
 \end{pmatrix}\,.\label{sl3rrank0}
 \ee
The matrix ${\cal M}$ is given by
 \be {\cal M} = \begin{pmatrix} - (1 + \rho\cos\beta - \ft18
 \rho^2\sin^2\beta) & -(\rho\cos\beta - \ft18\rho^2 \sin^2\beta) &
 \ft12\rho \sin\beta \cr -(\rho\cos\beta - \ft18\rho^2 \sin^2\beta)&
 (1 - \rho \cos\beta + \ft18 \rho^2 \sin^2\beta) & \ft12\rho
 \sin\beta\cr \ft12\rho \sin\beta & \ft12\rho \sin\beta & 1
 \end{pmatrix}\,.
 \ee

The classification of $\tilde C$ and $\tilde {\cal M}$, on the other
hand, is somewhat different than those discussed in \cite{wjl}.  We
may still use the rigid rescaling and the rigid gauge symmetry to make
 \be e^{\Phi}(\rho_0)=1\,,\qquad
\chi(\rho_0)\equiv A_{123}(\rho_0)=0 \ee
at any but a specific point $\rho_0$.  We shall choose it to be the
asymptotic infinity, which is located at $\rho=0$.  It follows that
 \be {\tilde{\cal M}(0)}={\rm diag}\{-1,1\}\,.
\ee
This ensures that the solutions are asymptotic flat.  We are now left
with an $SO(1,1)$ residual symmetry.  However, unlike the previous
$SL(3,\R)$ case, in which the $SL(3,\R)$ is part of the general coordinate
transformations in $D=11$, we have no reason to use this residual
symmetry further to put ${\tilde {\cal C}}$ into the simpler canonical
form obtained in \cite{wjl}.  Instead, we should regard this $SO(1,1)$
as a nontrivial solution-generating symmetry group and write down all of
the solutions.  In practice, it is convenient to use the canonical
forms obtained in \cite{wjl} and then recover the full sets of
inequivalent solutions by performing the $SO(1,1)$ transformations.
This leads to the following three classes of solutions.

{\bf Class i:} $\tilde{\cal C}$ has a pair of complex eigenvalues.
Its canonical form is
 \be {\tilde{\cal C}} =
 \begin{pmatrix} 0 & -\gamma \cr \gamma & 0 &
 \end{pmatrix}\,.
 \ee
The corresponding $\tilde {\cal M}$ is
 \be {\tilde{\cal M}} = \begin{pmatrix}-\cos{\gamma\rho} &
 \sin{\gamma\rho} &\cr \sin{\gamma\rho}& \cos{\gamma\rho}
 \end{pmatrix}\,.
 \ee
To restore the full $SO(1,1)$ multiplets, we perform a further $SO(1,1)$
transformation with
 \be {\tilde\Lambda} =
 \begin{pmatrix}
 \cosh{\theta} & \sinh{\theta} &\cr
 \sinh{\theta}& \cosh{\theta}
 \end{pmatrix}\,,
 \ee
where $\theta$ is an arbitrary constant.
The new $\tilde {\cal M}$ is then given by
 \be {\tilde{\cal M}} \rightarrow {\tilde\Lambda}^T{\tilde{\cal
 M}}{\tilde\Lambda}=\begin{pmatrix}-\cos{\gamma\rho}+
\sin{\gamma\rho}~\sinh{2\theta}
 & \sin{\gamma\rho}~\cosh{2\theta} \cr
 \sin{\gamma\rho}~\cosh{2\theta}&
 \cos{\gamma\rho}+\sin{\gamma\rho}~\sinh{2\theta}
 \end{pmatrix}\,.\ee
Thus we have
 \bea\label{m2_2_3.1}
 e^{-\Phi}&=&\cos{\gamma\rho}+\sin{\gamma\rho}~\sinh{2\theta}\cr
 \chi&=&\frac{\sin{\gamma\rho}~\cosh{2\theta}}{\cos{\gamma\rho}+
\sin{\gamma\rho}~\sinh{2\theta}}
 \eea

{\bf Class ii:} $\tilde{\cal C}$ has two real eigenvalues, with one
timelike and one spacelike eigenvectors. Its canonical form is
 \be {\tilde{\cal C}} = \begin{pmatrix} \gamma & 0
 \cr 0 & -\gamma
 \end{pmatrix}\,,
 \ee
and hence
 \be {\tilde{\cal M}} = \begin{pmatrix}-e^{\gamma\rho} & 0
 &\cr 0& e^{-\gamma\rho}
 \end{pmatrix}\,.
 \ee
A further $SO(1,1)$ transformation leads to
 \be {\tilde{\cal M}}
 \rightarrow {\tilde\Lambda}^T{\tilde{\cal
 M}}{\tilde\Lambda}=\begin{pmatrix}-(\cosh{\gamma\rho}+
\sinh{\gamma\rho}~\cosh{2\theta})
 & -\sinh{\gamma\rho}~\sinh{2\theta} &\cr
 -\sinh{\gamma\rho}~\sinh{2\theta}&
 \cosh{\gamma\rho}-\sinh{\gamma\rho}~\sinh{2\theta}
 \end{pmatrix}\,.\ee
Thus, we have
 \bea
 e^{-\Phi}&=&\cosh{\gamma\rho}-\sinh{\gamma\rho}~\cosh{2\theta}\cr
 \chi&=&-\frac{\sinh{\gamma\rho}~\sinh{2\theta}}{\cosh{\gamma\rho}
-\sinh{\gamma\rho}~\cosh{2\theta}}
 \eea

{\bf Class iii:} $\tilde{\cal C}$ is rank 1 and all of its eigenvalues are
zero. Its canonical form is
 \be {\tilde{\cal C}} = \begin{pmatrix} Q
 & Q \cr -Q & -Q
 \end{pmatrix}\,,
 \ee
which yields
 \be {\tilde{\cal M}} = \begin{pmatrix} -(1+Q\rho) &
 -Q\rho \cr -Q\rho& 1-Q\rho
 \end{pmatrix}\,.
 \ee
The further $SO(1,1)$ transformation leads to
 \be {\tilde{\cal M}}
 \rightarrow {\tilde\Lambda}^T{\tilde{\cal
 M}}{\tilde\Lambda}=\begin{pmatrix} -(1+e^{2\theta}Q\rho) &
 -e^{2\theta}Q\rho \cr -e^{2\theta}Q\rho& 1-e^{2\theta}Q\rho
 \end{pmatrix}\,.\ee
Thus we see that $\theta$ can be absorbed into the redefinition of
$Q$. Thus, in general we have
 \be e^{-\Phi}=1-Q\rho\,,\qquad \chi=-\frac{{Q\rho}}{1-Q\rho}\,. \ee

  We have thus obtained the solutions for both ${\cal M}$ and $\tilde
{\cal M}$.  The combination of the two gives rise to a total of 12
classes of solutions.  It is straightforward to lift these solutions
back to $D=11$ using the reduction ansatz discussed in section
(\ref{reductionansatz}).  These 12 classes of solutions comprise the
most general spherically symmetric M2-branes, with the isometry group
of $\R^{1,2}\times SO(8)$.

\subsection{Properties of the solutions}

In the previous subsection, we have shown that there exist twelve
different classes of spherically symmetric M2-branes. We shall now
discuss their properties.  To do this, we first make a coordinate
transformation so that the metric (\ref{ds8}) is manifestly conformal
flat, namely,
\bea \label{ds8plus}
 ds_8^2&=&g(\tilde r)(d{\tilde r}^2+{\tilde r}^2d\Omega_{7}^2)
=(1+\fft{a}{4{\tilde r}^{12}})^\fft13(d{\tilde r}^2+
{\tilde r}^2d\Omega_{7}^2)\,,
\eea
where
 \be {\tilde r}^6=\ft12(r^6+\sqrt{r^6-a}\,)\,.\ee
Thus we have
 \be \label{coord}
 r^{6}=\fft{4{\tilde r}^{12}+a}{4{\tilde r}^{6}}\,,\qquad
 f=\fft{(1-\fft{a}{ 4{\tilde r}^{12}})^2}{
   (1+\fft{a}{4{\tilde r}^{12}})^2}\,,\qquad
 \cos(6\sqrt{a}\rho)=
\fft{1-\fft{a}{4{\tilde r}^{12}}}{1+\fft{a}{4{\tilde r}^{12}}}\,.
 \ee
For $a\leq0$, there will be a naked curvature singularity at ${\tilde
r}^{12}=-{a/ 4}$ in general.  Note that the point where $e^{-\Phi}=0$
is also singular since the volume of the world volume becomes infinite
while that of the transverse space shrinks to zero.  To see this, let
us suppose that $e^{-\Phi}=0$ for a certain finite $\tilde r=\tilde r_0>0$, and
we find that both $\partial_{\rho}e^{-\Phi}$ and $\partial_{\tilde
r}{\rho}$ are finite at this point in all classes of the solutions. It
follows that the metric must be singular at this point, since
 \be e^{-\fft{\Phi}{3}}ds_8^2\sim (\tilde r-\tilde r_0)^\fft{1}{3}
(d\tilde r^2+\tilde r_0^2  d\Omega_{7}^2)\,.\ee
When such a singularity is avoided by an appropriate choice of the
parameters with a proper value of $\gamma$, we would have regular
solutions if ${\tilde r}^{12}=-{a/ 4}$ is a degenerate surface and
thus it is just a coordinate singularity. Such a solution is either a
black hole or a Kaluza-Klein bubble.

    For $a>0$, ${\tilde r}^{12}={a/ 4}$ could be just a coordinate
singularity and the metric can be regular if there are no further
singular terms from $e^\Phi$, which can be avoided if the point
$\coth{\gamma\rho}=-\cosh{2\theta}$ (or $\cot{\gamma\rho}=
\sinh{2\theta})$ is out of the range of our parameter except the
asymptotic point $\tilde r=0$.  In this case, the range of ${\tilde
  r}$ can be extended to $(0,\infty)$. Note that (\ref{ds8plus}) and
(\ref{coord}) are invariant under
 \be {\tilde r}^{6}\rightarrow \fft{a}{4{\tilde r}^{6}},\ee
and thus (\ref{ds8plus}) describes a wormhole connecting two asymptotic
regions $W_{0}$ and $W_{\infty}$ at $\td r=0$ and $\infty$
respectively. From now on, we shall denote $\td r$ by $r$ for
simplicity.

\bigskip\bigskip
{\bf Class I.i}
\bigskip

This class of solutions is the one that combines Class I for ${\cal
M}$ and Class i for $\tilde {\cal M}$.  Lifting the solutions back to
$D=11$, we have
 \bea
ds^2
 &=&e^{\fft{2}{3}\Phi}\left(e^{\a\rho}\cos{\b\rho}(dz_1^2-dt^2) +
2e^{\a\rho}\sin{\b\rho} dt dz_1+e^{-2\a\rho}dz_2^2\right)\cr
&&+e^{-\fft{1}{3}\Phi}(1+\ft{a}{4r^{12}})^\fft{1}{3}
(dr^2+r^2d\Omega_{7}^2)\,,\cr
A_\3&=& \fft{\sin{\gamma\rho}\cosh{2\theta}}{\cos{\gamma\rho}+
\sin{\gamma\rho}\sinh{2\theta}}dt\wedge dz^1\wedge dz^2\,,
\eea
where
 \bea
 && e^{-\Phi}=\cos{\gamma\rho}+\sin{\gamma\rho}\sinh{2\theta}\,,
\qquad \cos(6\sqrt{a}\rho)=\fft{1-\ft{a}{4r^{12}}}{1+\ft{a}{4r^{12}}}
\,,\cr
 && \b^2+\gamma^2-3\a^2=84a\,.\eea
If $a\leq0$, then $(-{a/ 4})^\fft{1}{12}< r<\infty$ corresponds to
$-\infty<\rho<0$. There is always a naked curvature singularity at
$\cot\gamma\rho=-\sinh{2\theta}$.  If $a>0$, the range for $r$ is $0<
r<\infty$, corresponding to $-\pi<6\sqrt{a}\rho<0$.  The point $r=({a/
4})^\fft{1}{12}$ corresponds to $6\sqrt{a}\rho=\pi/2$.  To avoid the
possible curvature singularity at $\cot{\gamma\rho}= -\sinh{2\theta}$,
we must have
 \bea\left\{
 \begin{array} {cc}
 -\fft{\pi\gamma}{6\sqrt{a}}\geq-{\rm arccot}(\sinh2\theta)
>-\pi {\rm ~~~~~for~~~} \gamma>0~,\\
 -\fft{\pi\gamma}{6\sqrt{a}}\leq{\rm arccot}(-\sinh2\theta)<\pi
{\rm ~~~~~for~~~}\gamma<0~.\end{array}\right.\label{Iigamma}\eea
The solutions then describe smooth M2-brane wormholes that connect
two asymptotic regions.  One asymptotic region lies at $r\rightarrow
\infty$, for which we have
 \be\rho\rightarrow-\fft{1}{6r^6}~,\qquad
e^{-\Phi}\rightarrow1-\fft{\gamma\sinh2\theta}{6r^6}\,.
\ee
It follows that
\bea
 ds^2&\rightarrow&\left(1+\left(\ft{2}{3}\gamma\sinh2\theta-\a\right)
\fft{1}{6r^6}\right)(-dt^2+dz_1^2)-\fft{2\beta}{6r^6}dtdz_1\cr
&& +\left(1+\left(\ft{2}{3}\gamma\sinh2\theta+2\a\right)
\fft{1}{6r^6}\right)dz_2^2 +(1-\fft{\gamma\sinh2\theta}{18r^6})
(dr^2+r^2d\Omega_{7}^2)\,,\cr
 A_\3&\rightarrow&-\fft{\gamma\cosh2\theta}{6r^6}dt\wedge
 dz^1\wedge dz^2\,. \eea
We shall call this asymptotic region $W_\infty$, which is the
eleven-dimensional Mankowski spacetime.  The ADM mass can be
calculated from the subleading terms of this asymptotic region, given
by
  \bea
  M=\int_{r\rightarrow\infty} d^7\Sigma^m(\partial_nh_{mn}-\partial_m
  h_{ii})=\Omega_7(\a-\gamma\sinh2\theta)\,,
  \eea
where
 \be\Omega_{p-1}=\fft{2\pi^\ft{p}{2}}{\Gamma(\ft{p}{2})}\ee
is the volume of the unit sphere $S^p$. The momentum along the $z_1$
direction is
 \be P_1=\beta\Omega_7\,.\ee
The M2-brane charge is
 \be Q_{M2}=\int_{r\rightarrow\infty}*F^{(4)}=\int_{r\rightarrow\infty}
d^7\Sigma^m F_{m012}=\gamma\cosh(2\theta)\, \Omega_7\,.\ee

     The other asymptotic region is at $r=0$, which we shall
denote as $W_0$.   When the inequality of (\ref{Iigamma}) is
strictly held, $W_0$ is an asymptotically flat region.  To see
this, let us define $R^{6}=a/4r^{6}$.   For $r\rightarrow 0$, we have
\bea
\rho&\rightarrow&\fft{1}{6R^6}-\fft{\pi}{6\sqrt a}~,\cr
 e^{-\Phi}&\rightarrow&\left[\cos{\left(\fft{\gamma\pi}{6\sqrt{a}}\right)}
-\sin{\left(\fft{\gamma\pi}{ 6\sqrt{a}}\right)}\sinh{2\theta}\right]
 \left[1+\frac{\sin{\left(\ft{\gamma\pi}{ 6\sqrt{a}}\right)}+
\cos{\left(\ft{\gamma\pi}{6\sqrt{a}}\right)}\sinh{2\theta}}
 {\cos{\left(\ft{\gamma\pi}{6\sqrt{a}}\right)}-
\sin{\left(\ft{\gamma\pi}{6\sqrt{a}}\right)}\sinh{2\theta}}
\fft{\gamma}{6R^6}\right]\cr
&&=A(1+\fft{\gamma B}{6R^6})\,,\eea
where
 \be
 A=\cos{(\fft{\gamma\pi}{6\sqrt{a}})}-
\sin{(\fft{\gamma\pi}{6\sqrt{a}})}\sinh{2\theta}~,\qquad
 B=\frac{\sin{(\fft{\gamma\pi}{6\sqrt{a}})}+
\cos{(\fft{\gamma\pi}{6\sqrt{a}})}\sinh{2\theta}}
 {\cos{(\fft{\gamma\pi}{6\sqrt{a}})}-
\sin{(\fft{\gamma\pi}{6\sqrt{a}})}\sinh{2\theta}}\,.
\ee
Making the follow coordinate redefinition
\bea
 \tilde z_1&=&A^{-\fft13}
  e^{-\ft{\a\pi}{12\sqrt{a}}}\left[z_1\cos{\left(\fft{\b\pi}{12\sqrt{a}}
\right)}-t\sin{\left(\fft{\b\pi}{12\sqrt{a}}\right)}\right]\,, \cr
 \tilde t&=&A^{-\fft13} e^{-\ft{\a\pi}{12\sqrt{a}}}
\left[z_1\sin{\left(\fft{\b\pi}{12\sqrt{a}}\right)}+
t\cos{\left(\fft{\b\pi}{12\sqrt{a}}\right)}\right]\,,\cr
\tilde z_2&=&A^{-\fft13} e^{\ft{\a\pi}{6\sqrt{a}}} z_2\,,
 \eea
we have
\bea
 ds^2&\rightarrow&\left[1-(\ft23\gamma B-\alpha)\fft{1}{6R^6}\right]
(-d\tilde t^2+d\tilde z_1^2) +\fft{2\beta}{6R^6}d\tilde td\tilde z_1
 +\left[1-(\ft23\gamma B+2\alpha)\fft{1}{6R^6}\right]d\tilde z_2^2
 \cr&&
 +(1+\fft{\gamma B}{18R^6})A^{\fft13}(dR^2+R^2d\Omega_{7}^2)\,,\cr
 A_\3&\rightarrow&-A^{-1}\cosh2\theta\sin{\left(\fft{\gamma \pi}{\sqrt{a}}
\right)}\left[1-\gamma\left(\cot{\left(\fft{\gamma \pi}{6\sqrt{a}}\right)}
+B\right)\fft{1}{6R^6}\right]dt\wedge dz^1\wedge dz^2\cr
 &=&-\cosh2\theta\sin{\left(\fft{\gamma \pi}{6\sqrt{a}}\right)}
\left[1-\fft{\gamma}{6A\,R^6\sin{(\ft{\gamma \pi}{6\sqrt{a}})}}
\right] d\tilde t\wedge d\tilde z^1\wedge d\tilde z^2\,.
\eea
The ADM mass, momentum, and M2 charges, measured in $W_0$,  are given by, respectively,
 \bea M=(\gamma B-\a)A\,\Omega_7,\qquad
P_1=-\b A\, \Omega_7\,,\qquad
Q_{M2}=-\gamma \cosh(2\theta)\, \Omega_7\,.
\eea

If, instead, the equality in (\ref{Iigamma}) is held, namely,
 \bea\left\{
 \begin{array} {cc}
 -\ft{\gamma\,\pi}{6\sqrt{a}}=-{\rm arccot}(\sinh2\theta)>-\pi
{\rm ~~~~~for~~~} \gamma>0~,\\
 -\ft{\gamma\, \pi}{6\sqrt{a}}={\rm arccot}(-\sinh2\theta)<\pi
{\rm ~~~~~for~~~}\gamma<0~,\end{array}\right.
\eea
the asymptotic region $W_0$ becomes AdS$_4\times S^7$.  This can be
seen easily since we have
 \bea
 e^{-\Phi}&\rightarrow& \gamma\sin{\left(\fft{\gamma\pi}{
6\sqrt{a}}\right)}\cosh^2{2\theta}\fft{1}{6R^6}\,.
 \eea

\bigskip\bigskip
{\bf Class I.ii}
\bigskip

The solution is
 \bea
ds^2 &=&e^{\fft23\Phi}(e^{\a\rho}\cos{\b\rho}(dz_1^2-dt^2)
+2e^{\a\rho}\sin{\b\rho} dt dz_1+e^{-2\a\rho}dz_2^2)\cr
&&+e^{-\fft13\Phi}(1+\fft{a}{4r^{12}})^\fft13(dr^2+r^2d\Omega_{7}^2)\,,\cr
A_\3&=&-\frac{\sinh{\gamma\rho}~\sinh{2\theta}}{
\cosh{\gamma\rho}-\sinh{\gamma\rho}~\cosh{2\theta}}
dt\wedge dz^1\wedge dz^2\,, \eea
where
 \bea
 && e^{-\Phi}=\cosh{\gamma\rho}-\sinh{\gamma\rho}~\cosh{2\theta}~,
\qquad \cos(6\sqrt{a}\rho)=\fft{1-\ft{a}{4r^{12}}}{1+\ft{a}{4r^{12}}}
~,\cr
&& \b^2-\gamma^2-3\a^2=84a~.
\eea
In the case of $a\leq0$, the coordinate $r$ lies in the range of
$(-{a/ 4})^\fft1{12}< r<\infty$, corresponding to $-\infty<\rho<0$.
For $\gamma<0$, there is curvature singularity at
$\coth\gamma\rho=\cosh{2\theta}$ , and the range of $\rho$ lies in
$\gamma^{-1}{\rm arccoth}(\cosh{2\theta})<\rho<0$.  For $\gamma\geq0$,
the range of $\rho$ is $-\infty<\rho<0$, and $r=(-{a/ 4})^\fft{1}{12}$
is a naked singularity in general, except for the case of the M2-brane
bubble, arising at
 \be\label{bubble1}\a=-\ft23\gamma~,\qquad
\fft{\gamma}{\sqrt{-a}}=6~,\qquad\beta=0\,.
\ee
The solution is a double Wick rotation of a black M2-brane \cite{hor},
analogous to the Kaluza-Klein bubble which is a double Wick rotation
of a black hole.

    In the case of $a>0$, we have $0< r<\infty$, corresponding to
$-\pi<6\sqrt{a}\rho<0$.  The point $r=({a/ 4})^\fft{1}{12}$
corresponds to $6\sqrt{a}\rho=\pi/2$.  To avoid the possible curvature
singularity at $\coth\gamma\rho=\cosh{2\theta}$, we must have
 \be\label{Iiigamma}
-\fft{\gamma\pi}{6\sqrt{a}}\leq{\rm arccoth}(\cosh2\theta)\,. \ee
The solutions describe wormholes that connect two asymptotic regions
$W_\infty$ and $W_0$ at $r=\infty$ and $r=0$, respectively.

      As $r\rightarrow\infty$, we have
 \be\rho\rightarrow-\fft{1}{6r^6}~,\qquad
e^{-\Phi}\rightarrow1+\fft{\gamma\cosh2\theta}{6r^6}\,,\ee
and then
\bea
 ds^2&\rightarrow&\left[1-\left(\ft23\gamma\cosh2\theta+\alpha\right)
\fft{1}{6r^6}\right](-dt^2+dz_1^2)-\fft{2\beta}{6r^6}dtdz_1\cr
&& +\left[1-\left(\ft23\gamma\cosh2\theta-2\a\right)\fft{1}{6r^6}
\right]dz_2^2
+(1+\fft{\gamma\cosh2\theta}{18r^6})(dr^2+r^2d\Omega_{7}^2)\,,\cr
 A_\3&\rightarrow&\fft{\gamma\sinh2\theta}{6r^6}dt\wedge dz^1\wedge dz^2
\,.
\eea
Thus the asymptotic region $W_\infty$ is flat.
The ADM mass, momentum, and M2 charges are given by, respectively,
 \be M=(\a+\gamma \cosh2\theta)\Omega_7~,\qquad
P_1=\beta\Omega_7~,\qquad Q_{M2}=-\gamma\sinh(2\theta)\Omega_7\,.\ee

     We now look at the asymptotic region $W_0$, which is
also flat if the inequality (\ref{Iiigamma}) is strictly held.
Let us define $R^{6}=a/4r^{6}$, and we have
 \bea\rho&\rightarrow&\fft{1}{6R^6}-\fft{\pi}{6\sqrt a}~,\cr
 e^{-\Phi}&\rightarrow&\left[\cosh{(\ft{\gamma\pi}{6\sqrt{a}})}+
\sinh{(\ft{\gamma\pi}{6\sqrt{a}})}\cosh{2\theta}\right]
 \left[1-\frac{\sinh{(\ft{\gamma\pi}{6\sqrt{a}})}+
\cosh{(\ft{\gamma\pi}{6\sqrt{a}})}\cosh{2\theta}}
 {\cosh{(\ft{\gamma\pi}{6\sqrt{a}})}+\sinh{(\ft{\gamma\pi}{6\sqrt{a}})}
\cosh{2\theta}}\fft{\gamma}{6R^6}\right]\cr
&&=A(1+\fft{\gamma B}{6R^6})\,,\eea
where
 \be
 A=\cosh{(\ft{\gamma\pi}{6\sqrt{a}})}+
\sinh{(\ft{\gamma\pi}{6\sqrt{a}})}\cosh{2\theta}~,\qquad
 B=-\frac{\sinh{(\ft{\gamma\pi}{6\sqrt{a}})}+
\cosh{(\ft{\gamma\pi}{6\sqrt{a}})}\cosh{2\theta}}
 {\cosh{(\ft{\gamma\pi}{6\sqrt{a}})}+
\sinh{(\ft{\gamma\pi}{6\sqrt{a}})}\sinh{2\theta}}\,.
\ee
Introducing new asymptotic coordinates
 \bea
 \tilde z_1&=&A^{-\fft13}
  e^{-\ft{\a\pi}{12\sqrt{a}}}\left[z_1\cos{(\ft{\b\pi}{12\sqrt{a}})}
-t\sin{(\ft{\b\pi}{12\sqrt{a}})}\right]\,, \cr
 \tilde t&=&A^{-\fft13}
  e^{-\ft{\a\pi}{12\sqrt{a}}}\left[z_1\sin{(\ft{\b\pi}{12\sqrt{a}})}
+ t\cos{(\ft{\b\pi}{12\sqrt{a}})}\right]\,,\cr
 \tilde z_2&=&A^{-\fft13} e^{\ft{\a\pi}{6\sqrt{a}}} z_2\,,
 \eea
we obtain
\bea
 ds^2&\rightarrow&\left[1-\left(\ft23\gamma B-\a\right)\fft{1}{6R^6}
\right](-d\tilde t^2+d\tilde z_1^2)
 +\fft{\beta}{3R^6}d\tilde td\tilde z_1
 +\left[1-\left(\ft23\gamma B+2\a\right)\fft{1}{6R^6}\right]
d\tilde z_2^2\cr&&
 +(1+ \fft{\gamma B}{18R^6})A^{\fft13}(dR^2+R^2d\Omega_{7}^2)\,,\cr
 A_\3&\rightarrow&A^{-1}\sinh2\theta\sinh{\left(
\fft{\gamma \pi}{6\sqrt{a}}\right)}
\left[1-\left(\coth{\left(\fft{\gamma \pi}{6\sqrt{a}}\right)}+B\right)
\fft{\gamma}{6R^6}\right]dt\wedge dz^1\wedge dz^2\cr
 &=&\sinh2\theta\sinh{\left(\fft{\gamma \pi}{6\sqrt{a}}\right)}
\left[1-\fft{\gamma}{6A R^6\sinh{(\ft{\gamma \pi}{6\sqrt{a}})}}
\right] d\tilde t\wedge d\tilde z^1\wedge d\tilde z^2\,.\eea
Indeed, the metric is flat when $R\rightarrow \infty$.
The ADM mass, momentum and M2 charges are given by, respectively,
 \be
M=(\gamma B-\a)A \Omega_7 ~,\qquad
P_1=-\b A \Omega_7\,,\qquad
Q_{M2}=\gamma \sinh(2\theta)\, \Omega_7\,.\ee

    If, on the other hand, the equality in (\ref{Iiigamma}) is held,
then $W_0$ is asymptotically AdS$_4\times S^7$.  This is because
 \be
 e^{-\Phi}\rightarrow \sinh{\left(\fft{\gamma\pi}{
6\sqrt{a}}\right)}\sinh^2{2\theta}\fft{\gamma}{6R^6}\,.
 \ee

\bigskip\bigskip
{\bf Class I.iii}
\bigskip

The solution is
\bea
ds^2&=&e^{\fft23\Phi} (e^{\a\rho}\cos{\b\rho}(dz_1^2-dt^2) +
2e^{\a\rho}\sin{\b\rho} dt dz_1+e^{-2\a\rho}dz_2^2)\cr
&&+e^{-\fft{1}{3}\Phi}(1+\fft{a}{4r^{12}})^\fft13
(dr^2+r^2d\Omega_{7}^2)\,,\cr
 A_\3&=&-\frac{{Q\rho}}{1-Q\rho}dt\wedge
  dz^1\wedge dz^2 \eea
where
 \be
e^{-\Phi}=1-Q\rho\,,\qquad
 \cos(6\sqrt{a}\rho)=\fft{1-\ft{a}{4r^{12}}}{1+\ft{a}{4r^{12}}}\,,
\qquad \b^2-3\a^2=84a\,.
\ee
In the case of $a\leq0$, we have $(-{a/ 4})^\fft{1}{12}< r<\infty$, and
correspondingly, the range of $\rho$ is $-\infty<\rho<0$.  For $Q<0$,
there is a curvature singularity at $Q\rho=1$ , and the range of $\rho$
is $Q^{-1}<\rho<0$. For $Q\geq0$, the range of $\rho$ is
$-\infty<\rho<0$, and $r=(-{a/ 4})^\fft{1}{12}$ is singular, except for
the extremal case $\alpha=\beta=0$.

In the case of $a>0$, we have $0< r<\infty$, and hence
$-\pi<6\sqrt{a}\rho<0$. The point of $r=({a/ 4})^\fft{1}{12}$
corresponds to $6\sqrt{a}\rho=\pi/2$ . To avoid the possible curvature
singularity associated with singular $e^{\Phi}$, we need
 \be\label{Iiiigamma}
 -\fft{Q\pi}{6\sqrt{a}}\leq1\,. \ee
The asymptotic region $W_\infty$ at $r\rightarrow\infty$ is flat.
To see this, we note that
 \be\rho\rightarrow-\fft{1}{6r^6}~,\qquad e^{-\Phi}
\rightarrow1+\fft{Q}{6r^6}\,.\ee
It follows that
\bea
 ds^2&\rightarrow&\left(1-\left(\ft23 Q+\a\right)\fft{1}{6r^6}\right)
(-dt^2+dz_1^2)-\fft{2\beta}{6r^6}dtdz_1
 +\left(1-(\ft23 Q-2\alpha)\fft{1}{6r^6}\right)dz_2^2\cr
&&+(1+\fft{Q}{18r^6})(dr^2+r^2d\Omega_{7}^2)\,,\cr
 A_\3&\rightarrow& \fft{Q}{6r^6}dt\wedge dz^1\wedge dz^2\,.
\eea
The ADM mass, momentum, and M2 charges are given by, respectively
 \be M=(\a+Q)\Omega_7~~,\qquad P_1=\beta \Omega_7~~,\qquad
Q_{M2}=-Q\Omega_7\,.\ee

     The asymptotic region $W_0$ at $r=0$ is also flat, provided that
the inequality of (\ref{Iiiigamma}) is strictly held.
Let us define $R^{6}=a/4r^{6}$, and then
 \bea\rho&\rightarrow& \fft{1}{6R^6}-\fft{\pi}{6\sqrt a}~,\cr
 e^{-\Phi}&\rightarrow&1+\fft{Q\pi}{6\sqrt a}-\fft{Q}{6R^6}
=A(1- \fft{Q}{6AR^6})\,,
\eea
where
 \be
 A=1+\fft{Q\pi}{6\sqrt a}\,. \ee
Further introducing new asymptotic coordinates
 \bea
 \tilde z_1&=&A^{-\fft13}
  e^{-\ft{\a\pi}{12\sqrt{a}}}\left(z_1\cos{(\ft{\b\pi}{12\sqrt{a}})}
-t\sin{(\ft{\b\pi}{12\sqrt{a}})}\right)\,, \cr
 \tilde t&=&A^{-\fft13}
  e^{-\ft{\a\pi}{12\sqrt{a}}}\left(z_1\sin{(\ft{\b\pi}{12\sqrt{a}})}+
t\cos{(\ft{\b\pi}{12\sqrt{a}})}\right)\,,\cr
 \tilde z_2&=&A^{-\fft13} e^{\ft{\a\pi}{6\sqrt{a}}} z_2\,,
 \eea
we have
\bea
 ds^2&\rightarrow&\left(1+(\ft{2Q}{3A}+\alpha)\fft{1}{6R^6}\right)
(-d\tilde t^2+d\tilde z_1^2) +\fft{2\beta}{6R^6}d\tilde td\tilde z_1
 +\left(1+(\ft{2Q}{3A}-2\a)\fft{1}{6R^6}\right)d\tilde z_2^2\cr
&& +(1-\fft{Q}{18A R^6})A^{\fft{1}{3}}(dR^2+R^2d\Omega_{7}^2)\,,\cr
 A_\3&\rightarrow& \fft{Q \pi}{6\sqrt{a}}
\left(1-\fft{\sqrt{a}}{\pi A R^6}\right) d\tilde t\wedge
 d\tilde z^1\wedge d\tilde z^2\,.\eea
The ADM mass, momentum, and M2 charges are given by, respectively,
 \bea M=-(Q+\a A)\Omega_7 ~~,\qquad
P_1=-\b A\Omega_7~~,\qquad Q_{M2}=Q\Omega_7\,.
\eea

   If on the other hand, the equality in (\ref{Iiiigamma}) is
held, we have
 \bea
 e^{-\Phi}&\rightarrow& -\fft{Q}{6R^6}\,.
 \eea
It follows that the asymptotic region $W_0$ is AdS$_4\times S^7$.
This solution was obtained and discussed in detail in \cite{adsworm}.

   Thus we see that all class I.i, class I.ii and class I.iii solutions
contain M2-brane wormholes that smoothly connect two asymptotic
regions.  One asymptotic region is always flat, while the other can
be either flat or AdS$_4\times S^7$.

\bigskip\bigskip
{\bf Class II.i}
\bigskip

The solution is
 \bea &&ds^2
 =e^{\fft23\Phi}\left(-e^{\a_1\rho}dt^2+ e^{\a_2\rho} dz_1^2
+e^{-(\a_1+\a_2)\rho}dz_2^2\right)+e^{-\fft13\Phi}
(1+\fft{a}{4r^{12}})^\fft13 (dr^2+r^2d\Omega_{7}^2)\,,\cr
 &&A_\3=\fft{\sin{\gamma\rho}\cosh{2\theta}}{\cos{\gamma\rho}+
\sin{\gamma\rho}\sinh{2\theta}}dt\wedge dz^1\wedge dz^2\,, \eea
where
 \bea
 && e^{-\Phi}=\cos{\gamma\rho}+\sin{\gamma\rho}\sinh{2\theta}~,\qquad
 \cos(6\sqrt{a}\rho)=\fft{1-\ft{a}{4r^{12}}}{1+\ft{a}{4r^{12}}}~,\cr
 && \gamma^2-\a_1^2-\a_2^2-\a_1\a_2=84a~.\eea
If $a\leq0$, $(-{a/ 4})^{\ft1{12}}< r<\infty$ corresponds to
$-\infty<\rho<0$. There is always a naked curvature singularity at
$\cot\gamma\rho=-\sinh{2\theta}$.  If $a>0$, $0< r<\infty$ corresponds
to $-\pi<6\sqrt{a}\rho<0$ and $r=({a/ 4})^\fft{1}{12}$ corresponds to
$6\sqrt{a}\rho=\pi/2$ . To avoid the possible curvature singularity at
$\cot{\gamma\rho}=-\sinh{2\theta}$, we must have
 \bea\left\{
 \begin{array} {cc}
 -\fft{\gamma\pi}{6\sqrt{a}}\geq-{\rm arccot}(\sinh2\theta)>
-\pi {\rm ~~~~~for~~~} \gamma>0~,\\
 -\fft{\gamma\pi}{6\sqrt{a}}\leq{\rm arccot}(-\sinh2\theta)<\pi
{\rm ~~~~~for~~~}\gamma<0~.\end{array}\right.\eea
However, from the constraint, we find that
$|\ft{\gamma}{\sqrt{a}}|>2\sqrt{21}>6$.  Thus the naked curvature
singularity at $\cot{\gamma\rho}=-\sinh{2\theta}$ is unavoidable, and
there is no regular solution in this class. The special case with
$\alpha_1=\alpha_2=0$ was obtained and discussed in detail in
\cite{lu&roy}.

\bigskip\bigskip
{\bf Class II.ii}
\bigskip

The solution is
\bea &&ds^2
 =e^{\fft23\Phi}\left(-e^{\a_1\rho}dt^2+ e^{\a_2\rho} dz_1^2 +
e^{-(\a_1+\a_2)\rho}dz_2^2\right)+e^{-\fft13\Phi}
(1+\fft{a}{4r^{12}})^\fft13(dr^2+r^2d\Omega_{7}^2)\,,\cr
 &&A_\3=-\frac{\sinh{\gamma\rho}~\sinh{2\theta}}{\cosh{\gamma\rho}
-\sinh{\gamma\rho}~\cosh{2\theta}}dt\wedge dz^1\wedge dz^2\,, \eea
where
 \bea
 && e^{-\Phi}=\cosh{\gamma\rho}-\sinh{\gamma\rho}~\cosh{2\theta}~,~~~~
 \cos(6\sqrt{a}\rho)=\fft{1-\ft{a}{4r^{12}}}{1+\ft{a}{4r^{12}}}~,\cr
 && -\gamma^2-\a_1^2-\a_2^2-\a_1\a_2=84a~.\eea
In this class we always have $a\leq0$. Then the range $(-{a/
  4})^\fft{1}{12}< r<\infty$ corresponds to $-\infty<\rho<0$. In the
case $\gamma<0$, there is a curvature singularity at
$\coth\gamma\rho=\cosh{2\theta}$, and the range of $\rho$ should be
$\gamma^{-1}{\rm arccoth}(\cosh{2\theta})<\rho<0$. In the case
$\gamma\geq0$, the range of $\rho$ should be $-\infty<\rho<0$, and
$r=(-{a/ 4})^\fft{1}{ 12}$ is singular in general, except in the cases
when the black M2-brane (or its double Wick rotation
counterpart M2-brane bubble (\ref{bubble1}) which is shared by classes I.ii and II.ii ) arises, namely,
 \be \a_1=-2\a_2~,~~~\a_2=-\ft23\gamma~,\qquad
\fft{\gamma}{\sqrt{-a}}=6\,.\ee
Another special case with $\alpha_1=\alpha_2=0$ was obtained and
discussed in detail in \cite{lpxtoda,lu&roy}.

\bigskip\bigskip
{\bf Class II.iii}
\bigskip

The solution is
 \bea &&ds^2
 =e^{\fft23\Phi}\left(-e^{\a_1\rho}dt^2+ e^{\a_2\rho} dz_1^2 +
e^{-(\a_1+\a_2)\rho}dz_2^2\right)+e^{-\fft13\Phi}
(1+\fft{a}{4r^{12}})^\fft13(dr^2+r^2d\Omega_{7}^2)\,,\cr
 &&A_\3=-\frac{{Q\rho}}{1-Q\rho}dt\wedge dz^1\wedge dz^2\,, \eea
where
 \bea
 && e^{-\Phi}=1-Q\rho~,\qquad
 \cos(6\sqrt{a}\rho)=\fft{1-\ft{a}{4r^{12}}}{1+\ft{a}{4r^{12}}}~,\cr
 && -\a_1^2-\a_2^2-\a_1\a_2=84a~.\eea
In this class we always have $a\leq0$. Then $(-{a/ 4})^\fft{1}{12}<
r<\infty$ corresponds to $-\infty<\rho<0$. In the case $Q<0$, there
is a curvature singularity at $Q\rho=1$ , and the range of $\rho$
should be $Q^{-1}<\rho<0$. In the case $Q\geq0$, the range of $\rho$
should be $-\infty<\rho<0$, and $r=(-{a/ 4})^\fft{1}{12}$ is the
singularity except for the extremal case $\a_1=\a_2=0$, which
corresponds to the BPS M2-brane \cite{dust}.

\bigskip\bigskip
{\bf Class III.i}
\bigskip

The solution is
 \bea ds^2&
 =&e^{\fft{2}{3}\Phi}\left\{e^{\a\rho}\left[-(1+\tilde Q\rho) dt^2-
2\tilde Q\rho dtdz_1+(1-\tilde Q\rho) dz_1^2 \right]
+e^{-2\a\rho}dz_2^2\right\}\cr&&+e^{-\fft13\Phi}
(1+\fft{a}{4r^{12}})^\fft13(dr^2+r^2d\Omega_{7}^2)\,,\cr
 A_\3&=&\fft{\sin{\gamma\rho}\cosh{2\theta}}{\cos{\gamma\rho}+
\sin{\gamma\rho}\sinh{2\theta}}dt\wedge dz^1\wedge dz^2\,, \eea
where
\bea
 && e^{-\Phi}=\cos{\gamma\rho}+\sin{\gamma\rho}\sinh{2\theta}~,\qquad
 \cos(6\sqrt{a}\rho)=\fft{1-\ft{a}{4r^{12}}}{1+\ft{a}{4r^{12}}}~,\cr
 && \gamma^2-3\a^2=84a~.\eea
If $a\leq0$, the coordinate range $(-{a/ 4})^\fft{1}{12}< r<\infty$
corresponds to $-\infty<\rho<0$. There should always be a curvature singularity
at $\cot\gamma\rho=-\sinh{2\theta}$ where $\rho$ is finite.  If $a>0$,
$0< r<\infty$ corresponds to $-\pi<6\sqrt{a}\rho<0$, and $r=({a/
  4})^\fft{1}{12}$ corresponds to $6\sqrt{a}\rho=\pi/2$ . To avoid the
possible curvature singularity at $\cot{\gamma\rho}=-\sinh{2\theta}$,
we need
 \bea\left\{
 \begin{array} {cc}
 -\fft{\gamma\pi}{6\sqrt{a}}\geq-{\rm arccot}(\sinh2\theta)
>-\pi {\rm ~~~~~for~~~} \gamma>0~,\\
 -\fft{\gamma\pi}{6\sqrt{a}}\leq{\rm arccot}
(-\sinh2\theta)<\pi {\rm ~~~~~for~~~}\gamma<0~.\end{array}\right.\eea
On the other hand, we find $|\fft{\gamma}{\sqrt{a}}| >2\sqrt{21}
>6$. Thus the curvature singularity at $\cot{\gamma\rho}=
-\sinh{2\theta}$ is unavoidable.

\bigskip\bigskip
{\bf Class III.ii}
\bigskip

The solution is
 \bea ds^2&
 =&e^{\fft{2}{3}\Phi}\left\{e^{\a\rho}\left[-(1+\tilde Q\rho) dt^2
-2\tilde Q\rho dtdz_1+(1-\tilde Q\rho) dz_1^2 \right]
+e^{-2\a\rho}dz_2^2\right\}\cr
&&+e^{-\fft13\Phi}(1+\fft{a}{4r^{12}})^\fft13(dr^2+r^2d\Omega_{7}^2)
\,,\cr
 A_\3&=&-\frac{\sinh{\gamma\rho}~\sinh{2\theta}}{\cosh{\gamma\rho}
-\sinh{\gamma\rho}~\cosh{2\theta}}dt\wedge dz^1\wedge dz^2\,, \eea
where
 \bea
 && e^{-\Phi}=\cosh{\gamma\rho}-\sinh{\gamma\rho}~\cosh{2\theta}~,~~~~
 \cos(6\sqrt{a}\rho)=\fft{1-\ft{a}{4r^{12}}}{1+\ft{a}{4r^{12}}}~,\cr
 && -\gamma^2-3\a^2=84a~.\eea
In this class we always have $a\leq0$. Then $(-{a/ 4})^\fft{1}{12}<
r<\infty$ corresponds to $-\infty<\rho<0$. In the case $\gamma<0$,
there is a curvature singularity at $\coth\gamma\rho=\cosh{2\theta}$,
and the range of $\rho$ should be $\gamma^{-1}{\rm
arccoth}(\cosh{2\theta})<\rho<0$. In the case $\gamma\geq0$, the range
of $\rho$ should be $-\infty<\rho<0$, and $r=(-{a/ 4})^\fft{1}{12}$ is
the singularity in general except for the M2-brane bubbles arising at
 \be \a=-\ft23\gamma~,~~~\fft{\gamma}{\sqrt{-a}}=6.\ee

\bigskip\bigskip
{\bf Class III.iii}
\bigskip

The solution is
 \bea ds^2&
 =&e^{\fft23\Phi}\left\{e^{\a\rho}\left[-(1+\tilde Q\rho)
dt^2-2\tilde Q\rho dtdz_1+(1-\tilde Q\rho)
 dz_1^2 \right]+e^{-2\a\rho}dz_2^2\right\}
 \cr&&+e^{-\fft13\Phi}(1+\fft{a}{4r^{12}})^\fft13
(dr^2+r^2d\Omega_{7}^2)\,,\cr
A_\3&=&-\frac{{Q\rho}}{1-Q\rho}dt\wedge dz^1\wedge dz^2\,, \eea
where
 \be
e^{-\Phi}=1-Q\rho\,,\qquad
 \cos(6\sqrt{a}\rho)=\fft{1-\ft{a}{4r^{12}}}{1+\ft{a}{4r^{12}}}
\,,\qquad -3\a^2=84a~.\ee
In this class we always have $a\leq0$. Then $(-{a/ 4})^\fft{1}{12}<
r<\infty$ corresponds to $-\infty<\rho<0$. In the case $Q<0$, there
is a curvature singularity at $Q\rho=1$ , and the range of $\rho$
should be $Q^{-1}<\rho<0$. In the case $Q\geq0$, the range of $\rho$
should be $-\infty<\rho<0$, and $r=(-{a/ 4})^\fft{1}{12}$ is the
singularity except for the extremal case $\a=0$.  In this limit, we have
$\rho\sim 1/r^{6}$, and the solution describes a pp-wave of momentum
$Q$ propagating in the worldvolume of a BPS M2-brane.

\bigskip\bigskip
{\bf Class IV.i}
\bigskip

The solution is
 \bea ds^2&
 =&e^{\fft23\Phi}\left[-\left(1+\a\rho\cos2\b-\ft18\a^2\rho^2\sin^2
 2\b\right)d t^2+\left(1-\a\rho\cos2\b+\ft18\a^2\rho^2\sin^2
 2\b\right)d z_1^2\right.\cr
&&\left.-2\left(\a\rho\cos2\b-
\ft18\a^2\rho^2\sin^2 2\b\right)dtdz_1+\a\rho\sin2\b dt d
 z_2+\a\rho\sin2\b d z_1d z_2+dz_2^2\right]\cr&&
+e^{-\fft13\Phi}(1+\fft{a}{4r^{12}})^\fft13(dr^2+r^2d\Omega_{7}^2)
\,,\cr
 A_\3&=&\frac{\sin{\gamma\rho}\cosh{2\theta}}{\cos{\gamma\rho}+
\sin{\gamma\rho}\sinh{2\theta}}dt\wedge dz^1\wedge dz^2\,,\eea
where
 \bea
 && e^{-\Phi}=\cos{\gamma\rho}+\sin{\gamma\rho}\sinh{2\theta}~,\qquad
 \cos(6\sqrt{a}\rho)=\fft{1-\ft{a}{4r^{12}}}{1+\ft{a}{4r^{12}}}~,\cr
 && \gamma^2=84a~.\eea
In this case we always have $a>0$. Then  $0< r<\infty$ corresponds
to $-\pi<6\sqrt{a}\rho<0$, and $r=({a/ 4})^\fft{1}{12}$ corresponds
to $6\sqrt{a}\rho=\pi/2$ . To avoid the possible curvature
singularity at $\cot{\gamma\rho}=-\sinh{2\theta}$, we need
 \bea\left\{
 \begin{array} {cc}
 -\fft{\gamma\pi}{6\sqrt{a}}\geq-{\rm arccot}(\sinh2\theta)
>-\pi {\rm ~~~~~for~~~} \gamma>0~,\\
 -\fft{\gamma\pi}{6\sqrt{a}}\leq{\rm arccot}(-\sinh2\theta)<\pi
{\rm ~~~~~for~~~}\gamma<0~.\end{array}\right.\eea
On the other hand, we find $|\ft{\gamma}{\sqrt{a}}|=2\sqrt{21}>6$. Thus the curvature singularity at
$\cot{\gamma\rho}=-\sinh{2\theta}$ is unavoidable, and there is no
regular solution in this class.

\bigskip\bigskip
{\bf Class IV.ii}
\bigskip

The solution is
 \bea ds^2&
 =&e^{\fft23\Phi}\left[-\left(1+\a\rho\cos2\b-\ft18\a^2\rho^2\sin^2
 2\b\right)d t^2+\left(1-\a\rho\cos2\b+\ft18\a^2\rho^2\sin^2
 2\b\right)d z_1^2\right.\cr
&&\left.-2\left(\a\rho\cos2\b-\ft18\a^2\rho^2\sin^2 2\b\right)
dtdz_1+\a\rho\sin2\b dt dz_2+\a\rho\sin2\b d z_1d z_2+dz_2^2\right]\cr
&&+e^{-\fft13\Phi}(1+\fft{a}{4r^{12}})^\fft13(dr^2+r^2d\Omega_{7}^2)\,,\cr
A_\3&=&-\frac{\sinh{\gamma\rho}~\sinh{2\theta}}{\cosh{\gamma\rho}
-\sinh{\gamma\rho}~\cosh{2\theta}}dt\wedge dz^1\wedge dz^2 \eea
where
\bea
 && e^{-\Phi}=\cosh{\gamma\rho}-\sinh{\gamma\rho}~\cosh{2\theta}~,
\qquad \cos(6\sqrt{a}\rho)=
\fft{1-\ft{a}{4r^{12}}}{1+\ft{a}{4r^{12}}}~,\cr
 && -\gamma^2=84a~.\eea
In this class we always have $a\leq0$. Then $(-{a/ 4})^\fft{1}{12}<
r<\infty$ corresponds to $-\infty<\rho<0$. In the case $\gamma<0$,
there is a curvature singularity at $\coth\gamma\rho=\cosh{2\theta}$,
and the range of $\rho$ should be $\gamma^{-1}{\rm
arccoth}(\cosh{2\theta})<\rho<0$. In the case $\gamma\geq0$, the
range of $\rho$ should be $-\infty<\rho<0$, and $r=(-{a/ 4})^\fft{1}{
12}$ is the singularity except the $\gamma=0$ limit which is the
tachyon wave, obtained in \cite{wjl}.

\bigskip\bigskip
{\bf Class IV.iii}
\bigskip

In this class we always have $a=0$. The solution is
 \bea ds^2&
 =&e^{\fft23\Phi}\left[-\left(1+\a\rho\cos2\b-\ft18\a^2\rho^2\sin^2
 2\b\right)d t^2+\left(1-\a\rho\cos2\b+\ft18\a^2\rho^2\sin^2
 2\b\right)d z_1^2\right.\cr
&&\left.-2\left(\a\rho\cos2\b-\ft18\a^2\rho^2\sin^2 2\b\right)dtdz_1
+\a\rho\sin2\b dt dz_2+\a\rho\sin2\b d z_1d z_2+dz_2^2\right]
 \cr&&+e^{-\ft13\Phi}(dr^2+r^2d\Omega_{7}^2)\,,\cr
 A_\3&=&-\frac{{Q\rho}}{1-Q\rho}dt\wedge dz^1\wedge dz^2\,, \eea
where
 \be e^{-\Phi}=1-Q\rho~,\qquad \rho=-\fft{1}{6r^{6}}.\ee
The range $0< r<\infty$ corresponds to $-\infty<\rho<0$. In the case
$Q<0$, there is a curvature singularity at $Q\rho=1$ , and the range of
$\rho$ should be $Q^{-1}<\rho<0$. In the case $Q\geq0$, the range of
$\rho$ should be $-\infty<\rho<0$, and $r=0$ is nonsingular.  This
configuration describes a tachyon wave propagating in the
M2-brane, obtained in \cite{wjl}.

\begin{table}[!h]
\caption{Solutions without a naked singularity} \vskip 0.2cm
\begin{tabular}{|c|c|c|}
\hline
Class   & Parameter regimes & Description  \\ \hline
I.i     & $a>0,~\gamma>0,~-\fft{\pi\gamma}{6\sqrt{a}}\geq-{\rm arccot}(\sinh2\theta)$ & Wormhole \\ \hline
I.i     & $a>0,~\gamma<0,~-\fft{\pi\gamma}{6\sqrt{a}}\leq{\rm arccot}(-\sinh2\theta)$ & Wormhole \\ \hline
I.ii    & $a>0,~-\fft{\pi\gamma}{6\sqrt{a}}\leq{\rm arccoth}(\cosh2\theta)$ & Wormhole \\ \hline
I.ii (also in II.ii)  & $a<0,~\gamma=6\sqrt{-a},~\a=-4\sqrt{-a},~\beta=0$ & M2 bubble\\ \hline
I.iii   & $a>0,~-\fft{\pi Q}{6\sqrt{a}}\leq1$ & Wormhole \\ \hline
I.iii (also in II.iii) & $a=\a=\b=0,~Q\geq0$ & BPS M2 brane \\ \hline
II.ii   & $a<0,~\gamma=6\sqrt{-a},~\a_2=-4\sqrt{-a},~\a_1=8\sqrt{-a}$ & Black M2 brane\\ \hline
III.ii  & $a<0,~\gamma=6\sqrt{-a},~\a=-4\sqrt{-a}$ &  M2 bubble with pp-wave \\ \hline
III.iii & $a=\a=0,~Q\geq0$ & pp-wave on BPS M2 brane \\ \hline
IV.iii  & $a=0,~Q\geq0$ & Tachyon wave on BPS M2 brane \\ \hline

\end{tabular}
\end{table}
In summary, we have obtained the most
general spherically symmetric M2-branes.
We found that there are a total of twelve classes
of solutions.  All of the solutions without a naked singularity are
listed in Table 1.
\section{The traversability of wormholes}

We have obtained the most general spherically symmetric M2-branes in
the previous sections. From Table 1,
we can find that the class I.i, I.ii,
and I.iii solutions with appropriate parameters describe smooth
wormholes that connect two asymptotic regions.  We have obtained their ADM
masses, linear momenta, as well as the M2 charges measured in each
asymptotic region.  In this section, we discuss the traversability of
these wormholes.  The traversability of higher-dimensional Ricci-flat
wormholes was discussed in \cite{lmflat}.

\subsection{Geodesic motion}

The metrics of the class I.i, I.ii, and I.iii solutions have the same
form.  The Lagrangian for geodesic motion is given by
\bea
{\cal L} &=& \ft12 g_{\mu\nu} \dot x^\mu \dot x^\nu \cr
 &=& \ft12e^{-{\fft13\Phi}}\left(1+\fft{a}{4r^{12}}\right)^\fft13
\left[ \dot r^2 +
 r^2  (\dot \theta_1^2 + \sin^2\theta_1\,\dot\theta_2^2+\dots+
\sin^2\theta_1\dots\sin^2\theta_{6}\,\dot\theta_7^2)\right] \cr
&&+\ft12 e^{\fft23\Phi-2\a\rho}\dot z_2^2+\ft12 e^{\fft23\Phi+\a\rho}
 \left[\cos{\b\rho}\, (\dot z_1^2-\dot t^2)+2\sin{\b\rho}\,
\dot t \dot z_1 \right]\,.
\eea
where a dot denotes a derivative with respect to the
proper time $\tau$.  The conserved quantities are
\bea
 J&=&e^{-{\fft13\Phi}}\left(1+\fft{a}{4r^{12}}\right)^\fft13 r^2
\sin^2\theta_1\dots\sin^2\theta_{6}\,\dot\theta_7\,,\cr
 E&=&e^{\fft23\Phi+\a\rho}(\cos{\b\rho}\,\dot t
-\sin{\b\rho}\,\dot z)\,,\cr
 p_1&=&e^{\fft23\Phi+\a\rho}(\cos{\b\rho}\,\dot z
+\sin{\b\rho}\,\dot t)\,,\cr
 p_2&=&e^{\ft23\Phi-2\a\rho}\dot z_2\,.\eea
By the $SO(8)$ symmetry, we can choose an appropriate coordinate system for any given geodesic such that
$\theta_i\equiv \ft12 \pi~(i=1,2,\dots,6)$.  The Lagrangian itself is a constant for
geodesic motions, {\it i.e.}, ${\cal L}=-\ft12\epsilon$, where
$\epsilon=1,0,-1$ for the timelike, null and spacelike geodesics
respectively.  It follows that we have
 \be e^{-{\fft13\Phi}}\left(1+\fft{a}{4r^{12}}\right)^\ft13\dot r^2
= -\epsilon- e^{{\fft13\Phi}}\left(1+\fft{a}{4r^{12}}\right)^{-\fft13}
\frac{J^2}{r^2 }- e^{-\fft23\Phi+2\a\rho}p_2^2  + V\,, \ee
where the potential is
 \bea V&=&e^{-\fft23\Phi-\a\rho}\left[(E^2-p_1^2)\,\cos{\b\rho}+2E
 p_1\,\sin{\b\rho} \right]\cr
 &=&e^{-\fft23\Phi-\a\rho}(E^2+p_1^2)\cos{(\b\rho-\lambda)}\,,\eea
where
 \be\lambda=\arcsin\left(\fft{2Ep_1}{E^2+p_1^2}\right)\,.\ee
The asymptotic regions $W_\infty$ and $W_0$ correspond to
$\rho=0$ and $\rho=-\ft{\pi}{6\sqrt{a}}$ respectively.

For the class I.i, we have $\gamma^2\ft{\pi^2}{36a}<\pi^2$, and thus
$\b^2\ft{\pi^2}{36a}\geq(84a-\gamma^2)\ft{\pi^2}{36a}>
\ft{4\pi^2}{3}$.  For the classes I.ii and I.iii, we have $\b^2
\ft{\pi^2}{36a} \geq 84a \ft{\pi^2}{36a}= \ft{7\pi^2}{3}$.
It follows that the range of the angle $\b\rho$ from $W_\infty$ to
$W_0$ is always larger than $\pi$.  Therefore $V$ must take some
negative values in some region of $r$, and the wormhole is not
traversable for both the timelike and null geodesics.

\subsection{Timelike trajectories}

We now consider traversable timelike trajectories.
The metric of the wormholes can be recast into
\bea
ds^2&=&e^{\fft23\Phi+\a\rho}(\sin\ft12{\b\rho}\,du +
\cos{\ft12\b\rho}\,dv)(\sin{\ft12\b\rho}\,dv -
\cos{\ft12\b\rho}\, du) + e^{\fft23\Phi-2\a\rho}dz_2^2\cr
&& +e^{-{\fft13\Phi}}\left(1+\fft{a}{4r^{12}}\right)^{\ft13}
\left[dr^2 + r^2 \,d\Omega_7^2\right]\,, \eea
where $(u,v)$ are light-cone coordinates, defined by
\be u=t-z_1\,,\,v=t+z_1~.\ee
To get the timelike trajectory from $W_+$ to $W_-$, let us
consider
\bea
u&=&\eta(\cot{\ft12(\b\rho+c)}+1)^2\cos{\ft12c}-
\eta(\cot{\ft12(\b\rho+c)}-1)^2\sin{\ft12c}+c_1\,,\cr
v&=&\eta(\cot{\ft12(\b\rho+c)}-1)^2\cos{\ft12c}+
\eta(\cot{\ft12(\b\rho+c)}+1)^2\sin{\ft12c}+c_2\,,\eea
where $\eta,\,c,\,c_1,\,c_2$ are arbitrary constants. We have
\bea\label{match}
&&\sin{\ft12\b\rho}\,u' +
\cos{\ft12\b\rho}\,v'=\cos{\ft12\b\rho}\,u' -
\sin{\ft12\b\rho}\,v'\cr
&&=-\frac{\eta\b}{\sin^3{\ft12(\b\rho+c)}}
\fft{d\rho}{dr}=\frac{\eta\b}{\sin^3{\ft12(\b\rho+c)}}
\fft{\sin{(6\sqrt{a}\rho)}}{\sqrt{a} r}\,,\eea
where a prime denotes a derivative with respect to $r$.  Then the
timelike condition
\bea e^{\ft23\Phi+\a\rho}(\sin{\ft12\b\rho}\,u' +
\cos{\ft12\b\rho}\,v')(\sin{\ft12\b\rho}\,v' -
\cos{\ft12\b\rho}\, u')  + e^{-{\ft13\Phi}}\left(1+\fft{a}{4r^{12}}
\right)^\ft13<0\eea
implies
\bea\label{ieq}
\frac{\eta^2\b^2\, e^{\Phi+\a\rho}}{\sin^6(\ft12(\b\rho+c))\,
 r^{14}\left(1+\ft{a}{4 r^{12}}\right)^{\fft73}}>1\,.\eea
Note that
\bea
\frac{\eta^2\b^2\, e^{\Phi+\a\rho}}{\sin^6(\ft12(\b\rho+c))\,
r^{14}\left(1+\ft{a}{4 r^{12}}\right)^{\fft73}}\geq
\frac{\eta^2\b^2\, e^{\Phi+\a\rho}}{r^{14}\left(1+\ft{a}{4 r^{12}}
\right)^{\fft73}} \geq 0\,,\eea
and $0$ is only possible for $r\rightarrow0,\infty$. From the
continuation of the expression of the left-hand side of (\ref{ieq}), its
minimum in the region $r>({a/4})^{1/12}$ must be greater than $0$  if
it is greater than $0$ at $W_{\infty}$; its minimum in the region $0<r<({a/4})^{1/12}$ must be also
greater than $0$ if it is greater
than $0$ at $W_{0}$. The most relevant example for the former case is
$c=0$ where
\be
\frac{\eta^2\b^2\, e^{\Phi+\a\rho}}{\sin^6({\ft12\b\rho})\, r^{14}
\left(1+\ft{a}{4 r^{12}}\right)^{\fft73}}\rightarrow\infty \,,\qquad
{\rm as} \qquad r\rightarrow\infty,\ee
while the most relevant example for the later case is
$c=\ft{\b\pi}{6\sqrt a}$ where
\be
\frac{\eta^2\b^2\, e^{\Phi+\a\rho}}{\sin^6({\ft12\b\rho}+
\ft{\b\pi}{12\sqrt a})\, r^{14}\left(1+\ft{a}{4 r^{12}}
\right)^{\fft73}}\rightarrow\infty \,,\qquad
{\rm as} \qquad r\rightarrow0\,.
\ee
Thus along the trajectories
\bea
\label{t1} {\rm T_{\infty}:}&& \left\{
\begin{array} {ll}
u=\eta_{\infty}(\cot{\ft12\b\rho}+1)^2+c_1,\\
v=\eta_{\infty}(\cot{\ft12\b\rho}-1)^2+c_2,\\
\end{array}\right.\\
\label{t0} {\rm T_{0}:}&& \left\{
\begin{array} {ll}
u=\eta_{0}(\cot({\ft12\b\rho}+\fft{\b\pi}{12\sqrt a})+1)^2
\cos\fft{\b\pi}{12\sqrt a}\\
\qquad\qquad\qquad-\eta_0(\cot({\ft12\b\rho}+
\fft{\b\pi}{12\sqrt a})-1)^2\sin\fft{\b\pi}{12\sqrt a}+d_1,\cr
v=\eta_{0}(\cot({\ft12\b\rho}+\fft{\b\pi}{12\sqrt a})-1)^2
\cos\fft{\b\pi}{12\sqrt a}\\
\qquad\qquad\qquad+\eta_0(\cot({\ft12\b\rho}+
\fft{\b\pi}{12\sqrt a})+1)^2\sin\fft{\b\pi}{12\sqrt a}+d_2.
\end{array}\right.\eea
with sufficiently large $\eta^2_{\infty,0}$, one could travel from
$W_{\infty,0}$ to the throat of wormhole $r=({a/4})^{\fft{1}{12}}$,
respectively. If the two trajectories could be matched in
$r=({a/4})^{\fft{1}{12}}$ smoothly, then one could travel from
$W_\infty$ to $W_0$.

For the trajectories (\ref{t1}) and (\ref{t0}) matching at
$r=({a/4})^{1/12}$, we need
 \bea c_1\!\!&=&\!\!d_1+\left(\cot(\ft{\b\pi}{24\sqrt
 a})+1\right)^2\eta_{0}\cos\ft{\b\pi}{12\sqrt
 a}-\left(\cot(\ft{\b\pi}{24\sqrt
 a})-1\right)^2\left(\eta_0\sin\ft{\b\pi}{12\sqrt
 a}+\eta_{\infty}\right)\,,\cr
 c_2\!\!&=&\!\!d_2+\left(\cot(\ft{\b\pi}{24\sqrt
 a})-1\right)^2\eta_{0}\cos\ft{\b\pi}{12\sqrt
 a}+\left(\cot(\ft{\b\pi}{24\sqrt
 a})+1\right)^2\left(\eta_0\sin\ft{\b\pi}{12\sqrt
 a}-\eta_{\infty}\right)\,.\eea
For the trajectories matching smoothly, we further require both
$\sin{\ft12\b\rho}\,u' + \cos{\ft12\b\rho}\,v'$ and
$\cos{\ft12\b\rho}\,u' - \sin{\ft12\b\rho}\,v'$ to match at
$r=({a/4})^{1/12}$. From (\ref{match}) we know that it is satisfied
when
\be \eta_{\infty}=-\eta_{0}.\ee

In conclusion, our example for timelike trajectories from $W_\infty$
to $W_0$ is
 \bea
 u&=& \left\{
\begin{array} {ll}
-\eta (\cot{\ft12\b\rho}+1)^2+\eta\left(\cot(\ft{\b\pi}{24\sqrt
 a})+1\right)^2\cos\ft{\b\pi}{12\sqrt
 a}\cr\qquad\qquad-\eta\left(\cot(\ft{\b\pi}{24\sqrt
 a})-1\right)^2\left(\sin\ft{\b\pi}{12\sqrt
 a}-1\right)+d_1\,, &\quad (r> ({\ft14a})^\fft1{12})
\,;\cr
\eta(\cot({\ft12\b\rho}+\fft{\b\pi}{12\sqrt a})+1)^2
\cos\fft{\b\pi}{12\sqrt a}\cr
\qquad\qquad-\eta(\cot({\ft12\b\rho}+
\fft{\b\pi}{12\sqrt a})-1)^2\sin\fft{\b\pi}{12\sqrt a}+d_1\,,
&\quad (r<({\ft14 a})^\fft{1}{12})\,;
\end{array}\right.\cr
v&=&\left\{
\begin{array} {ll} -\eta(\cot{\ft12\b\rho}-1)^2+
\eta\left(\cot(\ft{\b\pi}{24\sqrt
 a})-1\right)^2\cos\ft{\b\pi}{12\sqrt
 a}\cr
\qquad\qquad+\eta\left(\cot(\ft{\b\pi}{24\sqrt
 a})+1\right)^2\left(\sin\ft{\b\pi}{12\sqrt
 a}+1\right)+d_2\,,
&\quad
(r>({\ft14a})^\fft{1}{12})\,;\cr
\eta (\cot({\ft12\b\rho}+\fft{\b\pi}{12\sqrt a})-1)^2
\cos\fft{\b\pi}{12\sqrt a}\cr
\qquad\qquad +\eta
(\cot({\ft12\b\rho}+\fft{\b\pi}{12\sqrt
a})+1)^2\sin\fft{\b\pi}{12\sqrt a}+d_2\,, &\quad
(r<({\ft14a})^{\fft1{12}})\,,
\end{array}\right.\eea
where $\eta$ is large enough.

The Lagrangian for this trajectory
\bea 2{\cal L} &=&
e^{\fft23\Phi+\a\rho}(\sin{\ft12\b\rho}\,\dot u +
\cos{\ft12\b\rho}\,\dot v)(\sin{\ft12\b\rho}\,\dot v -
\cos{\ft12\b\rho}\, \dot u) + e^{-{\fft13\Phi}}\left(1+\fft{a}{4
r^{12}}\right)^\fft13\dot r^2\cr
&=& -1 \eea
implies
\bea\dot r=\left\{
\begin{array} {l}
-\left(\frac{\eta^2\b^2\, e^{\fft23\Phi+\a\rho}}{\sin^6(\ft12\b\rho)\,
r^{14}\left(1+\ft{a}{4 r^{12}}\right)^{{2}}}-
e^{-{\fft13\Phi}}\left(1+\fft{a}{4
r^{12}}\right)^\fft13\right)^{-\fft12}\,,
\cr \qquad\qquad\qquad\qquad\qquad (r>({\ft14a})^\fft{1}{12})\,;\cr
-\left(\frac{\eta^2\b^2\,e^{\fft23\Phi+\a\rho}}{\sin^6({\ft12\b\rho}+
\fft{\b\pi}{12\sqrt a})\, r^{14}\left(1+\fft{a}{4 r^{12}}
\right)^{{2}}}-e^{-{\fft13\Phi}}\left(1+\fft{a}{4
r^{12}}\right)^\fft13\right)^{-\fft12}\,, \cr
\qquad\qquad\qquad\qquad\qquad(r<({\ft14a})^\fft{1}{12})\,.
\end{array}\right.\eea
The radial velocity is zero at infinity and finite everywhere. Since
the velocity is analytic except $r=({a/4})^{1/12}$ and continuous
everywhere, the acceleration $\ddot r$ should be finite everywhere and
would not be continuous only at $r=({a/4})^{1/12}$.  As a check, let us
consider the simple case which is in class I.i with $\gamma=\a=0$.  In
this case, we have
\bea \ddot r&=&-\fft12{\dot r}^4
\left[\frac{\eta^2\b^2}{\sin^6(\ft12\b\rho+c)\,
r^{14}\left(1+\ft{a}{4 r^{12}}\right)^{{2}}}
\left(\fft{6a}{r^{13}}\left(1+\fft{a}{4 r^{12}}\right)^{-1}-\fft{14}{r}
\right.\right.\cr
&&\left.\left.\qquad\qquad\qquad\qquad+
\fft{3\beta\sin6\sqrt{a}\rho}{\sqrt{a}r}
\cot(\ft12\b\rho+c)\right)+\fft{a}{r^{13}}\left(1+\fft{a}{4 r^{12}}
\right)^{-\fft23}\right],\eea
where
\bea c=\left\{
\begin{array} {l}~0\,, ~~~\qquad(r>({\ft14a})^\fft{1}{12})\,;\cr
\fft{\b\pi}{12\sqrt a}\,,\qquad(r<({\ft14a})^\fft{1}{12})\,.
\end{array}\right.\eea
Note that $\rho=-\fft{\pi}{12\sqrt a}$ at $r=({a/4})^{1/12}$, and thus
$\ddot r$ has a jump at this point coming from the
$\cot(\ft12\b\rho+c)$ term. At any other point, the expression of
$\ddot r$ is regular.

The proper acceleration is given by \be A^\mu = \dot U^\mu +
\Gamma^\mu{}_{\nu\rho} U^\nu\, U^\rho\,, \ee where $U^\mu=\dot
x^\mu$. Its explicit expression is rather complicated. However, we can
choose a coordinate system where the relevant $\Gamma^u{}_{\nu\rho}$
are finite everywhere while $U^\mu$ is analytic except
$r=({a/4})^{1/12}$ and continues everywhere.  Therefore the proper
acceleration should also be finite everywhere and not continuous only
at $r=0$.  In conclusion, we have obtained a timelike trajectory
from $W_\infty$ to $W_0$ which is regular everywhere except
$r=({a/4})^{1/12}$ where the acceleration would have a jump.  Such a
trajectory is physically acceptable.

\section{General spherically symmetric type IIB strings}

The $\fft{SL(3,\R)}{SO(1,2)}\times \fft{SL(2,\R)}{SO(1,1)}$ scalar
coset, which we used to construct the most general spherically
symmetric M2-branes, can also arise from the dimension reduction of
type IIB supergravity on $\R^{1,1}$.  The bosonic type IIB action in the
Einstein frame is given by \cite{typeiib,bho}
 \bea S_{\rm IIB}&=&\int d^{10}x\sqrt{-G}\left(R-\ft12(\partial\phi)^2
 -\ft12 e^{2\phi}(\del\chi_0)^2-\ft1{12} e^{-\phi}H_\3^2-\ft1{12}
e^{\phi}{\hat F}_\3^2\right.\cr
&&\qquad\qquad\qquad \left. -\ft1{480} \hat F_\5^2\right)
-\ft12\int C_\4\wedge H_\3\wedge F_\3\,,
 \eea
where $H_\3 = dB_\2$, $F_\3=dC_\2$, $\hat F_\3=F_\3- \chi_0\wedge H_\3$ and $\hat F_\5=
dC_\4 - \ft12 C_\2\wedge H_\3 + \ft12 B_\2\wedge F_\3$.  In addition,
a self-duality condition $\hat F_\5={*\hat F_\5}$ should be imposed.  Since
we consider the reduction to eight dimensions
and collect the scalars only, the self-dual 5-form field strength
plays no role in our discussion.

         The reduction ansatz for the metric and
the two 2-form tensors is given by
 \bea
 ds^2_{10}&=&e^{-\fft{1}{2\sqrt3}{\tilde\Phi}}ds_8^2 +
e^{\fft{\sqrt3}{2}{\tilde\Phi}}dz^{\rm T} {\tilde{\cal M}}\, dz\cr
&=&e^{-\fft{1}{2\sqrt3}{\tilde\Phi}}ds_8^2+
 e^{\fft{\sqrt3}{2}{\tilde\Phi}}\left(
 e^{-\Phi}(dz+\chi dt)^2
 -e^{\Phi}dt^2\right)\,,\cr
C_\2&=& \chi_1 dt\wedge dz\,,\qquad
B_\2= \chi_2 dt\wedge dz\,.\label{typeiib}
 \eea
The scalars $\chi$ and $\Phi$, associated with $\tilde {\cal M}$,
 form a complex scalar that is an $\ft{SL(2,\R)}{SO(1,1)}$ coset.
The axionic scalars $\chi_0$, $\chi_1$ and $\chi_2$, the dilaton
$\phi$ and the breathing mode ${\tilde\Phi}$ form the
$\ft{SL(3,\R)}{SO(1,2)}$ scalar coset with
 \be\label{add_sl3r}
 \cal M=\begin{pmatrix}
 -e^{\ft{2}{\sqrt3}{\tilde\Phi}}+e^{{\phi}-\ft{1}{\sqrt3}{\tilde\Phi}}
\chi_{1}^2+e^{-{\phi}-\ft{1}{\sqrt3}{\tilde\Phi}}\chi_{2}^2
 &e^{{\phi}-\ft{1}{\sqrt3}{\tilde\Phi}}\chi_{1} &
 e^{{\phi}-\ft{1}{\sqrt3}{\tilde\Phi}}\chi_{0}\chi_{1}-e^{-{\phi}-
\ft{1}{\sqrt3}{\tilde\Phi}}\chi_{2}\\
 e^{{\phi}-\ft{1}{\sqrt3}{\tilde\Phi}}\chi_{1} & e^{{\phi}-
\ft{1}{\sqrt3}{\tilde\Phi}} & e^{{\phi}-\ft{1}{\sqrt3}{\tilde\Phi}}
\chi_{0} \\
e^{{\phi}-\ft{1}{\sqrt3}{\tilde\Phi}}\chi_{0}\chi_{1}-e^{-{\phi}-
\ft{1}{\sqrt3}{\tilde\Phi}}\chi_{2} & e^{{\phi}-
\ft{1}{\sqrt3}{\tilde\Phi}}\chi_{0}
 & e^{-{\phi}-\ft{1}{\sqrt3}{\tilde\Phi}}+e^{{\phi}-
\ft{1}{\sqrt3}{\tilde\Phi}}\chi_{0}^2
 \end{pmatrix}.
 \ee
Comparing with (\ref{M}), we find ${\phi}=\Phi_1$,
${\tilde\Phi}=-\Phi_2$, $\chi_0=\chi_{23}$, $\chi_1=\chi_{10}$ and
$\chi_2=-\chi_{20}$. These relationships can also be derived from
the T duality of the type IIA and type IIB strings. The Lagrangian
for the scalar sector in eight dimensions is given by
\bea
 {\cal L}&=&\sqrt{g} \Big[R_8 + \ft14 {\rm tr} (\partial_\mu {\cal
 M}^{-1}\partial^\mu{\cal M}) + \ft14 {\rm tr} (\partial_\mu {\tilde{\cal
 M}}^{-1}\partial^\mu{\tilde{\cal M}})\Big]\cr
 &=&\sqrt{g} \Big[R_8-\ft{1}{2}(\partial\Phi)^2
 +\ft12e^{-2\Phi}(\partial\chi)^2-\ft12(\partial{\phi})^2-
\ft12(\partial{\tilde\Phi})^2\cr
&&~~~~~~~~~~-\ft12e^{2{\phi}}(\partial\chi_{0})^2+
 \ft12e^{{\phi}-\sqrt3{\tilde\Phi}}(\partial\chi_{1}+
\chi_{2}\partial\chi_{0})^2 +\ft12e^{-{\phi}-\sqrt3{\tilde\Phi}}
(\partial\chi_{2})^2\Big]\,.
 \eea
Since the $SL(2,\R)$ symmetry associated with $\tilde {\cal M}$ is
part of the general coordinate transformation in type IIB supergravity, it
follows that we can put $\tilde{\cal M}$ in the simplified canonical
form.  Thus we have the followings:
\begin{itemize}

\item{\bf Class i:} $\tilde{\cal C}$ has a pair of complex eigenvalues,
 \be {\tilde{\cal M}} = \begin{pmatrix}-\cos{\gamma\rho} &
 \sin{\gamma\rho} &\cr \sin{\gamma\rho}& \cos{\gamma\rho}
 \end{pmatrix}\,.
 \ee

\item
{\bf Class ii:} $\tilde{\cal C}$ has two real
eigenvalues, with one timelike and one spacelike eigenvectors,
 \be {\tilde{\cal M}} = \begin{pmatrix}-e^{\gamma\rho} & 0
 &\cr 0& e^{-\gamma\rho}
 \end{pmatrix}\,.
 \ee

\item
{\bf Class iii:} $\tilde{\cal C}$ is rank 1 and all of its eigenvalues are
zero,
 \be {\tilde{\cal M}} = \begin{pmatrix} -(1+Q\rho) &
 -Q\rho \cr -Q\rho& 1-Q\rho
 \end{pmatrix}\,.
 \ee
\end{itemize}

For the $\ft{SL(3,\R)}{SO(1,2)}$ coset, the analysis is much more
complicated.  We can set ${\cal
M}(0)={\rm diag}\{-1,1,1\}$ by the rigid rescaling transformation and the
rigid gauge transformations. Then we are left with a rigid $SO(1,2)$
residual symmetry. The $SO(2)$ subgroup of $SO(1,2)$, which acts trivially
on $\tilde\Phi$, comes from the original classical $SL(2,\R)$ symmetry of
type IIB supergravity.

Given the solution of ${\cal M}$ in canonical form as in section 2.2,
the general solution can be obtained by performing the residual
$SO(1,2)$ transformation ${\cal M}\rightarrow{\Lambda}^T{{\cal
M}}{\Lambda}$. The $SO(1,2)$ transformation ${\Lambda}$ can be
parameterized by the analog of the Euler angles as
 \bea
 \Lambda&=&
 \begin{pmatrix}1 &
 0 & 0 \cr 0  & c_1 & -s_1\cr 0 &
 s_1 & c_1
 \end{pmatrix}\,
 \begin{pmatrix}ch &
 sh & 0 \cr sh  & ch & 0\cr 0
 & 0 & 1
 \end{pmatrix}\,
 \begin{pmatrix}1 &
 0 & 0 \cr 0  & c_2 & -s_2\cr 0 &s_2
 & c_2
 \end{pmatrix}\cr
&=&
 \begin{pmatrix}
 ch & c_2\, sh & - s_2\, sh \cr
c_1\, sh & \quad c_1 c_2\, ch - s_1 s_2 \quad
& -c_1 s_2\, ch - s_1 c_2\cr
s_1\, sh & \quad s_1 c_2\, ch + c_1 s_2 \quad &
-s_1 s_2\, ch + c_1 c_2 \end{pmatrix}\,,
 \eea
where $c_1=\cos\theta_1$, $c_2=\cos\theta_2$, $ch=\cosh\theta_0$,
$s_1=\sin\theta_1$, $s_2=\sin\theta_2$, and $sh=\sinh\theta_0$.

However, such parametrization gives
rise to lengthy and complicated results, so
we shall use an alternative parametrization as follows to
describe the most general solutions.

Given an arbitrary $\cal C$ matrix, we can transform it to be
the form
 \bea\label{add_toda} \cal C=\begin{pmatrix}
 \alpha_0    & -\beta_1   & -\beta_2  \cr%
 \beta_1     & \alpha_1    & 0         \cr%
 \beta_2     & 0      & \alpha_2        \cr%
 \end{pmatrix}~ \eea
by an $SO(2)$ transformation. Let us suppose that the eigenvalues
of $\cal C$ are $\lambda_{\mu}(\mu=0,1,2)$ and $\lambda_0$ is
related to the timelike or null eigenspace. We must have
 \be\det({\cal C}-\lambda I_{2p\times2p})=
\prod^2_{\mu=0}(\lambda_{\mu}-\lambda)\,.\ee
This implies
 \be \beta_i=\pm\sqrt{\frac{\prod^2_{\mu=0}
\left(\alpha_i-\lambda_{\mu}\right)}{\prod_{j\neq i}
(\alpha_i-\alpha_j)}}\,,\qquad i,j=1,2.
\ee

If $\cal C$ belongs to Class I or Class II, we can find a
transformation $T$ which transforms (\ref{add_toda}) to the canonical
form by solving the three eigenvectors of (\ref{add_toda}). Then
performing the inverse transformation $T^{-1}$ on the canonical
solution, we arrive at the solution for (\ref{add_toda}) as
 \bea\label{add_SOl-toda}
 {\cal M}=\Delta^{-1}\begin{pmatrix}
 \sum^2_{\mu=0}e^{\lambda_{\mu}\rho}\Lambda_{\mu}\a_{1,\mu}\a_{2,\mu}
& \b_1\sum^2_{\mu=0}e^{\lambda_{\mu}\rho}\Lambda_{\mu}\a_{2,\mu}   &
 \b_2\sum^2_{\mu=0}e^{\lambda_{\mu}\rho}\Lambda_{\mu}\a_{1,\mu}  \cr%
 \b_1\sum^2_{\mu=0}e^{\lambda_{\mu}\rho}\Lambda_{\mu}\a_{2,\mu}   &
\b^2_1\sum^2_{\mu=0}e^{\lambda_{\mu}\rho}\Lambda_{\mu}
\frac{\a_{2,\mu}}{\a_{1,\mu}}    &  \b_1\b_2\sum^2_{\mu=0}
e^{\lambda_{\mu}\rho}\Lambda_{\mu}         \cr%
 \b_2\sum^2_{\mu=0}e^{\lambda_{\mu}\rho}\Lambda_{\mu}\a_{1,\mu}
&\b_1\b_2\sum^2_{\mu=0}e^{\lambda_{\mu}\rho}\Lambda_{\mu}
& \b^2_2\sum^2_{\mu=0}e^{\lambda_{\mu}\rho}\Lambda_{\mu}
\frac{\a_{1,\mu}}{\a_{2,\mu}}       \cr%
 \end{pmatrix}. \eea
where
 \bea \Lambda_0&=&\lambda_1-\lambda_2,~~\Lambda_1=\lambda_2-\lambda_0,
~~\Lambda_2=\lambda_0-\lambda_1,\cr
\Delta&=&\prod^2_{\mu=0}\Lambda_{\mu},~~~ \a_{i,\mu}=
\a_i-\lambda_{\mu}.
 \eea
An additional $SO(2)$ transformation gives the parametrization of
general $\cal C$ as
 \bea\label{add_general3} \cal C=\begin{pmatrix}
 \alpha_0    & -(\beta_1\cos{\theta}-\beta_2\sin{\theta})
& -(\beta_2\cos{\theta}+\beta_1\sin{\theta})  \cr%
 \beta_1\cos{\theta}-\beta_2\sin{\theta}    & \a_1\cos^2{\theta}
+\a_2\sin^2{\theta}    & \ft12(\a_1-\a_2)\sin{2\theta} \cr%
\beta_2\cos{\theta}+\beta_1\sin{\theta}
&\ft12(\a_1-\a_2)\sin{2\theta}   & \a_1\sin^2{\theta}
+\a_2\cos^2{\theta}       \cr%
 \end{pmatrix}. \eea
The corresponding solutions in Class I and Class II are
 \bea\label{add_SOl-general3}
 {\cal M}=\Delta^{-1}\sum^2_{\mu=0}e^{\lambda_{\mu}\rho}\Lambda_{\mu}
\begin{pmatrix}
 \a_{1,\mu}\a_{2,\mu}    & B_{1,\mu}   & B_{2,\mu}  \cr%
 B_{1,\mu}   &  \frac{B_{1,\mu}^2}{\a_{1,\mu}\a_{2,\mu}}
&  \frac{B_{1,\mu}B_{2,\mu}}{\a_{1,\mu}\a_{2,\mu}}          \cr%
 B_{2,\mu}    &
 \frac{B_{1,\mu}B_{2,\mu}}{\a_{1,\mu}\a_{2,\mu}}&
\frac{B_{2,\mu}^2}{\a_{1,\mu}\a_{2,\mu}}       \cr%
 \end{pmatrix}~ ,\eea
where
 \be B_{1,\mu}=\b_1\a_{2,\mu}\cos{\theta}-\beta_2\a_{1,\mu}
\sin{\theta}\,, \qquad B_{2,\mu}=\b_2\a_{1,\mu}\cos{\theta}
+\beta_1\a_{2,\mu}\sin{\theta}\,.\ee
Comparing with (\ref{add_sl3r}), we can determine the solution of the
various fields to be
 \bea
 &&\chi_0=\frac{\sum^2_{\mu=0}e^{\lambda_{\mu}\rho}\Lambda_{\mu}
\frac{B_{1,\mu}B_{2,\mu}}{\a_{1,\mu}\a_{2,\mu}}}
 {\sum^2_{\mu=0}e^{\lambda_{\mu}\rho}\Lambda_{\mu}
\frac{B_{1,\mu}^2}{\a_{1,\mu}\a_{2,\mu}}},\,~~~~
\chi_1=\frac{\sum^2_{\mu=0}e^{\lambda_{\mu}\rho}
\Lambda_{\mu}B_{1,\mu}}{\sum^2_{\mu=0}e^{\lambda_{\mu}\rho}
\Lambda_{\mu}\frac{B_{1,\mu}^2}{\a_{1,\mu}\a_{2,\mu}}},\cr
&&e^{\phi-\ft{1}{\sqrt3}\tilde\Phi}=\Delta^{-1}{\sum^2_{\mu=0}
e^{\lambda_{\mu}\rho}\Lambda_{\mu}
\frac{B_{1,\mu}^2}{\a_{1,\mu}\a_{2,\mu}}},\cr
&& e^{-\ft{2}{\sqrt3}\tilde\Phi}=\Delta^{-2}\sum_{\mu>\nu}
e^{(\lambda_{\mu}+\lambda_{\nu})\rho}\Lambda_{\mu}\Lambda_{\nu}
\frac{\b_1^2\b_2^2(\a_{2,\mu}\a_{1,\nu}
-\a_{1,\mu}\a_{2,\nu})^2}{\a_{1,\mu}\a_{2,\mu}\a_{1,\nu}\a_{2,\nu}},\cr
&& \chi_2=\frac{\sum_{\mu>\nu}e^{(\lambda_{\mu}+\lambda_{\nu})\rho}
\Lambda_{\mu}\Lambda_{\nu} \frac{\b_1^2\b_2^2(\a_{2,\mu}\a_{1,\nu}-
\a_{1,\mu}\a_{2,\nu})(B_{1,\nu}\a_{1,\mu}\a_{2,\mu}-
B_{1,\mu}\a_{1,\nu}\a_{2,\nu})}{\a_{1,\mu}\a_{2,\mu}\a_{1,\nu}\a_{2,\nu}}}
 {\sum_{\mu>\nu}e^{(\lambda_{\mu}+\lambda_{\nu})\rho}
\Lambda_{\mu}\Lambda_{\nu}\frac{\b_1^2\b_2^2(\a_{2,\mu}\a_{1,\nu}
-\a_{1,\mu}\a_{2,\nu})^2}{\a_{1,\mu}\a_{2,\mu}\a_{1,\nu}\a_{2,\nu}}}.
 \eea
These expressions are still complicated and we shall
not discuss further about the properties of the solutions.

If the matrix $\cal C$ belongs to Class III, the twofold eigenvalues
$\lambda_0=\lambda_1=\lambda$ have only one eigenvector which is
null. We can solve the eigenvector $v_2$ for $\lambda_2$.  Based on
$v_2$, we can formulate an orthogonal and normalized basis that
transforms ${\cal M}$ into the canonical form. Performing the inverse
transformation on the canonical solution, we arrive at the solution
for (\ref{add_toda}) as
 \bea\label{add_SOl-toda3.1}
 {\cal M}&=&-\Lambda^{-2}\left[e^{\lambda\rho}
\begin{pmatrix}
 -C_0+\a_{1,1}\a_{2,1}\Lambda\rho    & \b_1(-{\a_{2,2}}
+\a_{2,1}\Lambda\rho)  &\b_2(-{\a_{1,2}}+\a_{1,1}\Lambda\rho) \cr%
 \b_1(-{\a_{2,2}}+\a_{2,1}\Lambda\rho)  &
\b_1^2\fft{-C_1+\a_{1,1}\a_{2,1}\Lambda\rho}{\a_{1,1}^2}  &
\b_1\b_2(-1+\Lambda\rho)          \cr%
 \b_2(-{\a_{1,2}}+\a_{1,1}\Lambda\rho)  &
 \b_1\b_2(-1+\Lambda\rho)       &
\b_2^2\fft{-C_2+\a_{1,1}\a_{2,1}\Lambda\rho}{\a_{2,1}^2}\cr%
 \end{pmatrix} \right.\cr
&&\left.~~~~~~~~~~~~~ +e^{\lambda_{2}\rho}
\begin{pmatrix}
 \a_{1,2}\a_{2,2}    & {\b_{1}}{\a_{2,2}}  & {\b_{2}}{\a_{1,2}}\cr%
 {\b_{1}}{\a_{2,2}}  &  {\b_{1}^2}\frac{\a_{2,2}}{\a_{1,2}}
&  \b_1\b_2   \cr%
 {\b_{2}}{\a_{1,2}} &\b_1\b_2       &  {\b_{2}^2}
\frac{\a_{1,2}}{\a_{2,2}}       \cr%
 \end{pmatrix}~ \right]\,,\eea
where
 \bea
 &&  \Lambda=\lambda-\lambda_2\,,\qquad
  \a_{i,j}=\a_i-\lambda_{j},\cr
 &&C_0=\a_{1,1}\a_{2,1}+\a_{1,1}\Lambda+\a_{2,1}\Lambda,\cr
 &&C_1=\a_{1,1}\a_{2,1}+\a_{1,1}\Lambda-\a_{2,1}\Lambda,\cr
 &&C_2=\a_{1,1}\a_{2,1}-\a_{1,1}\Lambda+\a_{2,1}\Lambda.
 \eea
This result can also be obtained as the $\lambda_0-\lambda_1\rightarrow0$
limit of (\ref{add_SOl-toda}). An additional $SO(2)$ transformation gives
the parametrization of general $\cal C$ as (\ref{add_general3}),
and the corresponding solution in Class III is
{\fontsize{10 pt}{\baselineskip}\selectfont
\bea\label{add_SOl-general3.1}
&&{\cal M}=-\Lambda^{-2}\left[ e^{\lambda_{2}\rho}\begin{pmatrix}
 \a_{1,2}\a_{2,2}    & B_{1,2}   & B_{2,2}  \cr%
 B_{1,2}   &  \frac{B_{1,2}^2}{\a_{1,2}\a_{2,2}}&
\frac{B_{1,2}B_{2,2}}{\a_{1,2}\a_{2,2}}          \cr%
B_{2,2} & \frac{B_{1,2}B_{2,2}}{\a_{1,2}\a_{2,2}}
&  \frac{B_{2,2}^2}{\a_{1,2}\a_{2,2}}       \cr%
 \end{pmatrix}\right.\\
&&\left.
 + e^{\lambda\rho}\begin{pmatrix}
 \Lambda^2-\a_{1,2}\a_{2,2}    & -B_{1,2}   & -B_{2,2}  \cr%
 -B_{1,2}   &  -\Lambda^2-\frac{B_{1,2}^2}{\a_{1,2}\a_{2,2}}
&  -\frac{B_{1,2}B_{2,2}}{\a_{1,2}\a_{2,2}}          \cr%
-B_{2,2} & -\frac{B_{1,2}B_{2,2}}{\a_{1,2}\a_{2,2}}
&  -\Lambda^2-\frac{B_{2,2}^2}{\a_{1,2}\a_{2,2}}       \cr%
 \end{pmatrix}+e^{\lambda\rho}\Lambda\rho\begin{pmatrix}
 \a_{1,1}\a_{2,1}    & B_{1,1}   & B_{2,1}  \cr%
 B_{1,1}   &  \frac{B_{1,1}^2}{\a_{1,1}\a_{2,1}}
&  \frac{B_{1,1}B_{2,1}}{\a_{1,1}\a_{2,1}}          \cr%
 B_{2,1}  & \frac{B_{1,1}B_{2,1}}{\a_{1,1}\a_{2,1}}
&  \frac{B_{2,1}^2}{\a_{1,1}\a_{2,1}}       \cr%
 \end{pmatrix} \right]\,,\nn\eea}
where
 \be
B_{1,i}=\b_1\a_{2,i}\cos{\theta}-\beta_2\a_{1,i}\sin{\theta}\,,
\qquad B_{2,i}=\b_2\a_{1,i}\cos{\theta}+\beta_1\a_{2,i}\sin{\theta}.
 \ee
Comparing with (\ref{add_sl3r}), we can decide the solution of the
various fields as
 \bea &&
 \chi_0=\frac{e^{\lambda\rho}(\frac{B_{1,2}B_{2,2}}{\a_{1,2}\a_{2,2}
\Lambda^2}-\rho\frac{B_{1,1}B_{2,1}}{\a_{1,1}\a_{2,1}\Lambda})
 -e^{\lambda_2\rho}\frac{B_{1,2}B_{2,2}}{\a_{1,2}\a_{2,2}\Lambda^2}}
 {e^{\lambda\rho}(1+\frac{B_{1,2}^2}{\a_{1,2}\a_{2,2}\Lambda^2}
 -\rho\frac{B_{1,1}^2}{\a_{1,1}\a_{2,1}\Lambda})-e^{\lambda_2\rho}
\frac{B_{1,2}^2}{\a_{1,2}\a_{2,2}\Lambda^2}},\cr
&& \chi_1=\frac{e^{\lambda\rho}(\frac{B_{1,2}}{\Lambda^2}-
\rho\frac{B_{1,1}}{\Lambda})-e^{\lambda_2\rho}\frac{B_{1,2}}{\Lambda^2}}
 {e^{\lambda\rho}(1+\frac{B_{1,2}^2}{\a_{1,2}\a_{2,2}\Lambda^2}
 -\rho\frac{B_{1,1}^2}{\a_{1,1}\a_{2,1}\Lambda})-e^{\lambda_2\rho}
\frac{B_{1,2}^2}{\a_{1,2}\a_{2,2}\Lambda^2}},\cr
&& e^{\phi-\ft13\tilde\Phi}=e^{\lambda\rho}(1+
\frac{B_{1,2}^2}{\a_{1,2}\a_{2,2}\Lambda^2}
-\rho\frac{B_{1,1}^2}{\a_{1,1}\a_{2,1}\Lambda})-e^{\lambda_2\rho}
\frac{B_{1,2}^2}{\a_{1,2}\a_{2,2}\Lambda^2},\cr
&& e^{-\fft{2}{\sqrt3}\tilde\Phi}=e^{(\lambda+\lambda_2)\rho}\left(
-\frac{\b_1^2\a_{2,2}^2+\b_2^2\a_{1,2}^2}{\a_{1,2}\a_{2,2}\Lambda^2}
 +\rho\frac{\b_1^2\b_2^2(\a_{1,2}\a_{2,1}-\a_{1,1}\a_{2,2})^2
}{\a_{1,1}\a_{1,2}\a_{2,1}\a_{2,2}\Lambda^3}\right)\cr
&&~~~~+e^{2\lambda\rho}\left(1+\frac{\b_1^2\a_{2,2}^2+
\b_2^2\a_{1,2}^2}{\a_{1,2}\a_{2,2}\Lambda^2}\right.\cr
&&\qquad\qquad\left.
 -\rho\frac{\b_1^2\b_2^2(\a_{1,2}\a_{2,1}-\a_{1,1}\a_{2,2})^2
+\a_{1,2}\a_{2,2}(\b_1^2\a_{2,1}^2+\b_2^2\a_{1,1}^2)}
 {\a_{1,1}\a_{1,2}\a_{2,1}\a_{2,2}\Lambda^3}\right),\\
&&
 e^{-\ft{2}{\sqrt3}\tilde\Phi}\chi_2=e^{(\lambda+\lambda_2)\rho}\left(
\frac{B_{2,2}}{\Lambda^2}-\rho\frac{\b_1^2\b_2^2(\a_{1,2}\a_{2,1}
-\a_{1,1}\a_{2,2}) (B_{1,1}\a_{1,2}\a_{2,2}-
B_{1,2}\a_{1,1}\a_{2,1})}{\a_{1,1}\a_{1,2}\a_{2,1}\a_{2,2}
\Lambda^3}\right)\cr
&&~~~+e^{2\lambda\rho}\left[-\frac{B_{2,2}}{\Lambda^2}
 +\rho\left(\frac{\b_1^2\b_2^2(\a_{1,2}\a_{2,1}-\a_{1,1}\a_{2,2})
 (B_{1,1}\a_{1,2}\a_{2,2}-B_{1,2}\a_{1,1}\a_{2,1})}{\a_{1,1}
\a_{1,2}\a_{2,1}\a_{2,2}\Lambda^3}
+\frac{B_{1,1}}{\Lambda^2}\right)\right].\nn\eea
Note that the $SL(2,\Z)$ multiplet of type IIB superstrings obtained
in \cite{sl2z} is a special case of the class III solutions.

If $\cal C$ belongs to Class IV, the canonical form given by
(\ref{sl3rrank0}) itself is related to the general form
(\ref{add_toda}) by an $SO(2)$ transformation.  Then the most general
solution for ${\cal C}$ for this class can be obtained from an
arbitrary $SO(2)$ transformation acting on the canonical ${\cal C}$
given in (\ref{sl3rrank0}).  Thus we have
\be {\fontsize{9 pt}{\baselineskip}\selectfont
\cal C= \a\begin{pmatrix}
 \cos{\b} & \ft14(3\cos(\beta-\theta)+\cos(\beta+\theta)) &
\,\,-\ft14(3\sin(\beta-\theta)-\sin(\beta+\theta)) \cr
  -\ft14(3\cos(\beta-\theta)+\cos(\beta+\theta))
& -\cos(\beta-\theta)\cos\theta & \ft12\sin(\beta-2\theta) \cr
\ft14(3\sin(\beta-\theta)-\sin(\beta+\theta))  &
\ft12\sin(\beta-2\theta) &  \sin(\beta-\theta)\sin\theta
 \end{pmatrix}}.\ee
Indeed, when $\theta=\ft12\beta$, the matrix takes the form of
(\ref{add_toda}).  The ${\cal M}$ is correspondingly given by
{\fontsize{10 pt}{\baselineskip}\selectfont
 \bea &&{\cal M} =\begin{pmatrix}
 -1 & 0 & 0 \cr
 0  & 1 & 0 \cr
 0  & 0 & 1
 \end{pmatrix}+\ft18\a^2\rho^2\sin^2{\b}
  \begin{pmatrix}
 1 &  \cos \theta  &  \sin \theta  \cr
  \cos \theta   &  \cos^2 \theta &
 \ft12\sin2 \theta\cr
  \sin \theta   &
 \ft12 \sin2 \theta &
 \sin^2 \theta
 \end{pmatrix}
 +\a\rho \times\\
&& \begin{pmatrix}
 \cos{\b} & -\ft14(3\cos(\beta-\theta)+\cos(\beta+\theta))
&\ft14(3\sin(\beta-\theta)-\sin(\beta+\theta)) \cr
  -\ft14(3\cos(\beta-\theta)+\cos(\beta+\theta))  &
-\cos(\beta-\theta)\cos\theta & \ft12\sin(\beta-2\theta) \cr
\ft14(3\sin(\beta-\theta)-\sin(\beta+\theta))  &
\ft12\sin(\beta-2\theta) &  \sin(\beta-\theta)\sin\theta
 \end{pmatrix}.\nn
 \eea}
Comparing with (\ref{add_sl3r}), we can determine that the scalar
fields are given by
  \bea
 \chi_0&=&\frac{\fft12\a\rho\sin(\beta-2\theta)+
\fft1{16}\a^2\rho^2\sin^2\beta\sin2\theta}
 {1-\a\rho\cos(\beta-\theta)\cos\theta+\fft18\a^2\rho^2\sin^2\beta
\cos^2\theta},\cr
\chi_1&=&\frac{-\fft14\a\rho(3\cos(\beta-\theta)+\cos(\beta+\theta))+
\fft18\a^2\rho^2\sin^2\beta\cos\theta}
 {1-\a\rho\cos(\beta-\theta)\cos\theta+\ft18\a^2\rho^2\sin^2\beta
\cos^2\theta},\cr
\chi_2&=&-\frac{\fft14\a\rho(3\sin(\beta-\theta)-
\sin(\beta+\theta))-\ft18\a^2\rho^2\sin^2\beta\sin\theta}
 {1-\a\rho\cos\beta-\ft18\a^2\rho^2\sin^2\beta},\cr
e^{\phi-\ft{1}{\sqrt3}\tilde\Phi}&=&1-\a\rho\cos(\beta-\theta)
\cos\theta+\ft18\a^2\rho^2\sin^2\beta\cos^2\theta,\cr
e^{-\fft{2}{\sqrt3}\tilde\Phi}&=&1-\a\rho\cos\beta-
\ft18\a^2\rho^2\sin^2\beta.
\eea

Having obtained the results for both $\tilde {\cal M}$ and
${\cal M}$ for all classes of solutions, it is straightforward to
lift the solution back to $D=10$ using (\ref{typeiib}), giving
rise to the most general string solutions with $\R^{1,1}\times SO(8)$
isometries.

\section{$D=5$ pure gravity on $\R^{1,1}$}

In this section, we consider five-dimensional pure gravity on
$\R^{1,1}$. The resulting $D=3$ gravity theory is described by an
$\ft{SL(3,\R)}{SO(1,2)}$ coset.  This enables us to find the most
general Ricci-flat solutions with $R^{1,1}\times SO(3)$ isometry.  The
Einstein-Hilbert action in five dimensions is given by
 \be
 S_5=\int d^{5}x\sqrt{-G}\, R\,.
 \ee
Performing the dimension reduction on $\R^{1,1}$, we can take
 \be
 ds^2_{5}=e^{-\ft{2}{\sqrt3}{\Phi}}ds_3^2+
 e^{\ft{1}{\sqrt3}{\Phi}}\left[
 -e^{\phi}(dt+\chi dz+\tilde A_\1)^2
 +e^{-\phi}(dz+ A_\1)^2\right]\,.
 \ee
The Lagrangian becomes
 \be
 {\cal L}_3=\sqrt{g}\Big(R_3-\ft12(\partial\phi)^2-\ft12(\partial\Phi)^2
 +\ft12 e^{2\phi}(\partial\chi)^2+\ft14 e^{\phi+\sqrt{3}\Phi}
\tilde F_\2^2-\ft14 e^{-\phi+\sqrt{3}\Phi}F_\2^2\Big)\,,
 \ee
where
 \be
 \tilde F_\2=d\tilde A_\1-d\chi\wedge A_\1\,,\qquad
 F_\2=dA_\1\,.
 \ee
The equations of motion are
 \bea
\fft{\d S}{\d\tilde A_\1}&=&-d{*\left(e^{\phi+\sqrt{3}\Phi}
\tilde F_\2\right)}=0~,\cr
 \fft{\d S}{\d A_\1}&=& d{*\left(e^{-\phi+\sqrt{3}\Phi}
F_\2\right)}-{*\left(e^{\phi+\sqrt{3}\Phi}\tilde
F_\2\right)}\wedge d\chi \cr
&=&d\left(e^{-\phi+\sqrt{3}\Phi}{*F_\2}-e^{\phi+\sqrt{3}\Phi}
{*\tilde F_\2}\wedge\chi\right)=0\,.
 \eea
Thus we can define the dual field of $\tilde A^{(1)}$ and
$A^{(1)}$ as follows
 \be d\tilde\varphi=-e^{\phi+\sqrt{3}\Phi}{*\tilde F_\2},\qquad
 d\varphi=e^{-\phi+\sqrt{3}\Phi}{*F_\2}-
e^{\phi+\sqrt{3}\Phi}{*\tilde F_\2\chi}~.\ee
Then we have
 \be \tilde F_\2={**\tilde F_\2}=-e^{-\phi-\sqrt{3}\Phi}
{*d\tilde\varphi}~,~~F_\2={**F_\2}=e^{\phi-\sqrt{3}\Phi}
{*(d\varphi-\chi\,d\tilde\varphi)}. \ee
The Lagrangian then describe an $\ft{SL(3,R)}{SO(1,2)}$ coset
 \bea
 {\cal L}&=&\sqrt{g}\Big[R_3-\ft12(\partial\phi)^2-\ft12(\partial\Phi)^2
 +\ft12e^{2\phi}(\partial\chi)^2-\ft12e^{-\phi-\sqrt{3}\Phi}
(\partial\tilde\varphi)^2
 +\ft12e^{\phi-\sqrt{3}\Phi}(\partial\varphi-
\chi\partial\tilde\varphi)^2\Big]\cr
&=&\sqrt{g}\Big[R_3+ \ft14 {\rm tr} (\partial_\mu
{\cal M}^{-1}\partial^\mu{\cal M})\Big]\,,
 \eea
where $\cal M$ is the metric related to the $\ft{SL(3,\R)}{SO(1,2)}$
scalar coset
 \bea\label{add_5DM}
 \cal M=\begin{pmatrix}
 -e^{-\phi+\ft{1}{\sqrt3}\Phi}+e^{\phi+\ft{1}{\sqrt3}\Phi}\chi^2
+e^{-\ft{2}{\sqrt3}\Phi}\varphi^2 &\quad
 e^{\phi+\ft{1}{\sqrt3}\Phi}\chi+e^{-\ft{2}{\sqrt3}\Phi}
\tilde\varphi\varphi \quad &
 e^{-\ft{2}{\sqrt3}\Phi}\varphi     \\
 e^{\phi+\ft{1}{\sqrt3}\Phi}\chi+e^{-\ft{2}{\sqrt3}\Phi}
\tilde\varphi\varphi &
 e^{\phi+\ft{1}{\sqrt3}\Phi}+e^{-\ft{2}{\sqrt3}\Phi}\tilde\varphi^2 &
 e^{-\ft{2}{\sqrt3}\Phi}\tilde\varphi  \\
 e^{-\ft{2}{\sqrt3}\Phi}\varphi & e^{-\ft{2}{\sqrt3}\Phi}\tilde\varphi
& e^{-\ft{2}{\sqrt3}\Phi}
 \end{pmatrix}.
 \eea

We shall consider the spherically symmetry solutions. As in section 2, the
metric in three dimensions is determined by the Einstein equation in
the foliating sphere directions; it is given by
 \be  ds_3^2=(1+\ft{a}{4{ r}^{2}})^{2}(dr^2+r^2d\Omega_{2}^2)\,.\ee
All of the scalars depend on the radial coordinate $r$ only.  The scalar
equations of motion can be integrated to satisfy the following
first-order equation:
 \be
{\cal M}^{-1} \dot {\cal M} = {\cal C}\,,
\ee
where the constant matrix ${\cal C}$ satisfies a constraint
from the Einstein equation associated with $R_{rr}$, given by
\be {\cal I}\equiv -\ft12 {\rm tr} ({\cal
 C}^2)= 4a\,.\ee
For this $\ft{SL(3,\R)}{SO(1,2)}$ coset, we can set ${\cal M}(0)={\rm
  diag}\{-1,1,1\}$ by a rigid coordinate and gauge transformation.  We
are then left with a rigid $SO(1,2)$ residual symmetry where there is
a $SO(1,1)$ subgroup coming from the rigid coordinate
transformation. Thus the nontrivial transformations of the solutions
form an $\ft{SO(1,2)}{SO(1,1)}$ coset.  Before modding out the trivial
$SO(1,1)$ part, the most general solutions for $\cal{C}$ and $\cal{M}$
take the same form as those obtained in previous section, but with
 \be\cos(\sqrt a\rho)=\frac{1-\ft{a}{4r^2}}{1+\ft{a}{4r^2}}\,.\ee
Comparing with (\ref{add_5DM}), we can find the solutions of the various
fields for different classes.

\bigskip
{\bf Class I \& II}:
\bigskip
\bea
&&\tilde\varphi=\frac{\sum^2_{\mu=0}e^{\lambda_{\mu}\rho}\Lambda_{\mu}
\frac{B_{1,\mu}B_{2,\mu}}{\a_{1,\mu}\a_{2,\mu}}}
{\sum^2_{\mu=0}e^{\lambda_{\mu}\rho}\Lambda_{\mu}
\frac{B_{2,\mu}^2}{\a_{1,\mu}\a_{2,\mu}}},\,~~~~
\varphi=\frac{\sum^2_{\mu=0}e^{\lambda_{\mu}\rho}
\Lambda_{\mu}B_{2,\mu}}
 {\sum^2_{\mu=0}e^{\lambda_{\mu}\rho}\Lambda_{\mu}
\frac{B_{2,\mu}^2}{\a_{1,\mu}\a_{2,\mu}}},\cr
&& e^{-\fft{2}{\sqrt3}\Phi}=\Delta^{-1}{\sum^2_{\mu=0}
e^{\lambda_{\mu}\rho}\Lambda_{\mu}
\frac{B_{2,\mu}^2}{\a_{1,\mu}\a_{2,\mu}}},\cr
&& e^{\phi-\fft{1}{\sqrt3}\Phi}=\Delta^{-2}\sum_{\mu>\nu}
e^{(\lambda_{\mu}+\lambda_{\nu})\rho}\Lambda_{\mu}\Lambda_{\nu}
 \frac{\b_1^2\b_2^2(\a_{2,\mu}\a_{1,\nu}-\a_{1,\mu}
\a_{2,\nu})^2}{\a_{1,\mu}\a_{2,\mu}\a_{1,\nu}\a_{2,\nu}},\cr
&& \chi= \frac{\sum_{\mu>\nu}e^{(\lambda_{\mu}+\lambda_{\nu})\rho}
\Lambda_{\mu}\Lambda_{\nu}\frac{\b_1^2\b_2^2(\a_{2,\mu}
\a_{1,\nu}-\a_{1,\mu}\a_{2,\nu})(B_{2,\nu}\a_{1,\mu}
\a_{2,\mu}-B_{2,\mu}\a_{1,\nu}\a_{2,\nu})}{\a_{1,\mu}
\a_{2,\mu}\a_{1,\nu}\a_{2,\nu}}}{\sum_{\mu>\nu}
e^{(\lambda_{\mu}+\lambda_{\nu})\rho}\Lambda_{\mu}\Lambda_{\nu}
\frac{\b_1^2\b_2^2(\a_{2,\mu}\a_{1,\nu}
-\a_{1,\mu}\a_{2,\nu})^2}{\a_{1,\mu}\a_{2,\mu}
\a_{1,\nu}\a_{2,\nu}}}.\eea

{\bf Class III:}
\bigskip
\bea &&
 \tilde\varphi=\frac{e^{\lambda\rho}(\frac{B_{1,2}B_{2,2}}{\a_{1,2}
\a_{2,2}\Lambda^2}-\rho\frac{B_{1,1}B_{2,1}}{
\a_{1,1}\a_{2,1}\Lambda}) -e^{\lambda_2\rho}\frac{B_{1,2}B_{2,2}}{
\a_{1,2}\a_{2,2}\Lambda^2}}
{e^{\lambda\rho}(1+\frac{B_{2,2}^2}{\a_{1,2}\a_{2,2}\Lambda^2}
 -\rho\frac{B_{2,1}^2}{\a_{1,1}\a_{2,1}\Lambda})-e^{\lambda_2\rho}
\frac{B_{2,2}^2}{\a_{1,2}\a_{2,2}\Lambda^2}},\cr
&&
 \varphi=\frac{e^{\lambda\rho}(\frac{B_{2,2}}{\Lambda^2}-
\rho\frac{B_{2,1}}{\Lambda})-e^{\lambda_2\rho}\frac{B_{2,2}}{\Lambda^2}}
 {e^{\lambda\rho}(1+\frac{B_{2,2}^2}{\a_{1,2}\a_{2,2}\Lambda^2}
 -\rho\frac{B_{2,1}^2}{\a_{1,1}\a_{2,1}\Lambda})-
e^{\lambda_2\rho}\frac{B_{2,2}^2}{\a_{1,2}\a_{2,2}\Lambda^2}},
 \cr&& e^{-\fft{2}{\sqrt3}\Phi}=e^{\lambda\rho}(1+
\frac{B_{2,2}^2}{\a_{1,2}\a_{2,2}\Lambda^2}
 -\rho\frac{B_{2,1}^2}{\a_{1,1}\a_{2,1}\Lambda})-
e^{\lambda_2\rho}\frac{B_{2,2}^2}{\a_{1,2}\a_{2,2}\Lambda^2},\cr
&&
e^{\phi-\fft{1}{\sqrt3}\Phi}=e^{(\lambda+\lambda_2)\rho}\left(-
\frac{\b_1^2\a_{2,2}^2+\b_2^2\a_{1,2}^2}{\a_{1,2}\a_{2,2}\Lambda^2}
 +\rho\frac{\b_1^2\b_2^2(\a_{1,2}\a_{2,1}-\a_{1,1}
\a_{2,2})^2}{\a_{1,1}\a_{1,2}\a_{2,1}\a_{2,2}\Lambda^3}\right)\cr
&&
+e^{2\lambda\rho}\left(1+\frac{\b_1^2\a_{2,2}^2+\b_2^2\a_{1,2}^2}{
\a_{1,2}\a_{2,2}\Lambda^2} -\rho\frac{\b_1^2\b_2^2(\a_{1,2}\a_{2,1}
-\a_{1,1}\a_{2,2})^2+\a_{1,2}\a_{2,2}(\b_1^2\a_{2,1}^2
+\b_2^2\a_{1,1}^2)}{\a_{1,1}\a_{1,2}\a_{2,1}\a_{2,2}\Lambda^3}\right),
\cr
&&
 e^{\phi-\fft{1}{\sqrt3}\Phi}\chi=e^{(\lambda+\lambda_2)\rho}\left(
-\frac{B_{1,2}}{\Lambda^2} -\rho
\frac{\b_1^2\b_2^2(\a_{1,2}\a_{2,1}-\a_{1,1}\a_{2,2})
 (B_{2,1}\a_{1,2}\a_{2,2}-B_{2,2}\a_{1,1}\a_{2,1})}{
\a_{1,1}\a_{1,2}\a_{2,1}\a_{2,2}\Lambda^3}\right)\cr
&&
+e^{2\lambda\rho}\left[\frac{B_{1,2}}{\Lambda^2}
 +\rho\left(\frac{\b_1^2\b_2^2(\a_{1,2}\a_{2,1}-\a_{1,1}\a_{2,2})
 (B_{2,1}\a_{1,2}\a_{2,2}-B_{2,2}\a_{1,1}\a_{2,1})}{
\a_{1,1}\a_{1,2}\a_{2,1}\a_{2,2}\Lambda^3}-
\frac{B_{1,1}}{\Lambda^2}\right)\right].\eea

{\bf Class IV:}
\bigskip
\bea &&
 \tilde\varphi=\frac{\ft12\a\rho\sin(\beta-2\theta)+\ft1{16}
\a^2\rho^2\sin^2\beta\sin2\theta}
 {1+\a\rho\sin(\beta-\theta)\sin\theta+\ft18
\a^2\rho^2\sin^2\beta\sin^2\theta},\cr
&& \varphi=\frac{\frac{1}{4}\a\rho(3\sin(\beta-\theta)-
\sin(\beta+\theta))+\ft18\a^2\rho^2\sin^2\beta\sin\theta}
 {1+\a\rho\sin(\beta-\theta)\sin\theta+\ft18
\a^2\rho^2\sin^2\beta\sin^2\theta},
 \cr&& e^{-\fft{2}{\sqrt3}\Phi}=1+\a\rho\sin(\beta-\theta)\sin\theta
+\ft18\a^2\rho^2\sin^2\beta\sin^2\theta,
 \cr&& e^{\phi-\fft{1}{\sqrt3}\Phi}=1-\a\rho\cos\beta-
\ft18\a^2\rho^2\sin^2\beta,
 \cr&& \chi= -\frac{\ft{1}{4}\a\rho(3\cos(\beta-\theta)
+\cos(\beta+\theta))+\ft18\a^2\rho^2\sin^2\beta\cos\theta}
 {1-\a\rho\cos\beta-\ft18\a^2\rho^2\sin^2\beta}.\eea
To write the 5D metric explicitly, we further need to convert the
solution of $\varphi$ and $\tilde\varphi$ to gauge potentials.
In fact, the gauge potentials $A_\1$ and $\tilde A_\1$ can
be decided without knowing the explicit form of $\varphi$ and
$\tilde\varphi$. Note that the components of ${\cal{M}}^{-1}\dot
{\cal{M}}={\cal{C}}$
\bea &&
e^{\phi-\sqrt3\Phi}(-\dot\varphi+\chi\dot{\tilde\varphi})=
{\cal{C}}_{02},\cr
&&
e^{-\phi-\sqrt3\Phi}\dot{\tilde\varphi}+\chi
e^{\phi-\sqrt3\Phi}(\dot\varphi-\chi\dot{\tilde\varphi})=
{\cal{C}}_{12},
\eea
imply
\bea &&
 dA_\1=F_\2=e^{\phi-\sqrt3\Phi}*(\dot\varphi-\chi
\dot{\tilde\varphi})dr =-{\cal{C}}_{02}d\Omega_2~~, \cr
&&d\tilde A_\1-d\chi\wedge A_\1-\chi dA_\1=\tilde F_\2
-\chi F_\2\cr
&&=-e^{-\phi-\sqrt3\Phi}*\dot{\tilde\varphi}dr-\chi
e^{\phi-\sqrt3\Phi}*(\dot\varphi-\chi \dot{\tilde\varphi})dr
 =-{\cal{C}}_{12}d\Omega_2~~.\eea
Then we can determine the gauge potential via ${\cal C}$ directly
\bea
&&A_\1={\cal{C}}_{02}\cos\theta_1d\theta_2~~,\cr&&
 \tilde A_\1={\cal{C}}_{12}\cos\theta_1d\theta_2+\chi A_\1
=({\cal{C}}_{12}+\chi{\cal{C}}_{02}) \cos\theta_1d\theta_2~~,\eea
where $\theta_1$ and $\theta_2$ are the angular coordinates for
the $S^2$ metric $d\Omega_2^2$.

Finally, let us consider how to mod out the trivial $SO(1,1)$.
The $SO(1,1)$ symmetry allows us to set the two constant and thus the
$U(1)$ potentials in the following canonical form:
\bea
a)~~{\cal{C}}_{02}&=&0; \cr b)~~{\cal{C}}_{12}&=&0; \cr
c)~~{\cal{C}}_{02}&=&{\cal{C}}_{12}=1.\eea
Applying this condition
to the solutions obtained previously, we then arrive with a full
classification of the solutions with the trivial $SO(1,1)$ being modded
out.

\section{Conclusions}

   In this paper we obtain the most general spherically symmetric
M2-brane and type IIB string solutions.  We make use of the fact that
any such $p$-brane solution reduced on the world volume give rise to
an instanton solution supported by a scalar coset in the lower-dimensional theory.  The Kaluza-Klein reduction of eleven-dimensional
supergravity on $\R^{1,2}$ or type IIB supergravity on $\R^{1,1}$ gives
rise to a scalar coset $\fft{SL(3,\R)}{SO(1,2)}\times \fft{SL(2,\R)}{
  SO(1,1)}$.  Using the classifications of $GL(n, \R)$ instantons for
the Ricci-flat examples \cite{wjl}, we obtain the most general
spherically symmetric instanton solutions.  Lifting these solutions
back to $D=11$ gives rise to M2-branes.

   We find that there are a total twelve classes of M2-branes, including
the previously known extremal or nonextremal M2-branes and new smooth
M2-brane wormholes that connect two asymptotic regions: one is flat
and the other can be either flat or AdS$_4\times S^7$.  We discuss
their properties in some detail.  We also obtained the most general
spherically symmetric type IIB string solutions by lifting the
instanton solutions.  Owing to the fact that the $SL(3,\R)$ and
$SL(2,\R)$ factors of the global symmetries are the general coordinate
transformation invariance in M theory and type IIB supergravity
respectively, the two types of lifting are very different, since we
would like to mod out solutions that are related by general coordinate
transformation.  We also obtain the most general Ricci-flat solutions
in five dimensions with $\R^{1,1}\times SO(3)$ isometries.

  As was discussed in \cite{lmp}, analytical continuation of $p$-brane
solutions can lead to cosmological solutions \cite{lmpx,low} where the
coordinate $r$ is analytically continued to become the time
coordinate.  These solutions can be interpreted as spacelike branes
\cite{gs}.  Thus a proper analytical continuation of our solution will
yield the most general S-branes with $\R^{p+1}\times SO(D-p-1)$
isometries,

   Our procedure can be generalized to obtain the most general solutions
with $\R^{1,p}\times SO(D-p-1)$ isometries in any supergravity in $D$
dimensions.  It amounts to a group theoretic study of a certain scalar
coset $G/H$ and then classifying the inequivalent classes of the constant
matrix ${\cal C}$ defined in (\ref{scalareom}).  The examples
discussed so far all belong to $GL(n,\R)$ or $SL(n,\R)$, for which the
discussion is relatively easy.  The situation becomes much more
complicated when $G$ belongs to exceptional groups, which can arise in
lower-dimensional maximal supergravities.  It is of great interest to
obtain all of the most general spherically symmetric $p$-branes in all
supergravities.

\section*{Acknowledgement}

Z.L.W. acknowledges the support by grants from the Chinese Academy of
Sciences, a grant from 973 Program with grant No: 2007CB815401 and
grants from the NSF of China with Grant No:10588503 and 10535060.

\end{document}